\documentclass[12pt]{iopart}

\usepackage[T1]{fontenc}
\usepackage{cite}
\usepackage{array}
\usepackage{placeins}
\usepackage{xcolor}
\usepackage{fancyhdr}
\usepackage{iopams}  
\usepackage{graphicx}
\usepackage[breaklinks=true,colorlinks=true,linkcolor=blue,urlcolor=blue,citecolor=blue]{hyperref}

\begin{document}

\title[Apparent anomalous diffusion and non-Gaussian distributions]
{Apparent anomalous diffusion and non-Gaussian distributions
in a simple mobile--immobile transport model with Poissonian switching}

\author{Timo J. Doerries$^{\dagger}$, Aleksei V. Chechkin$^{\dagger,
\ddagger,\S}$ and Ralf Metzler$^{\dagger}$}
\address{$\dagger$Institute of Physics \& Astronomy, University of Potsdam,
14476 Potsdam, Germany\\
$\ddagger$Faculty of Pure and Applied Mathematics, Hugo Steinhaus Center,
Wroc\l{}aw University of Science and Technology, Wyspianskiego 27,
50-370 Wroc\l{}aw, Poland\\
$\S$Akhiezer Institute for Theoretical Physics, 61108 Kharkov, Ukraine}

\begin{abstract}
We analyse mobile-immobile transport of particles that switch between
the mobile and immobile phases with finite rates. Despite this seemingly
simple assumption of Poissonian switching we unveil a rich transport
dynamics including significant transient anomalous diffusion and non-Gaussian
displacement distributions.  Our discussion is based on experimental parameters
for tau proteins in neuronal cells, but the results obtained here are expected to
be of relevance for a broad class of processes in complex systems. Concretely,
we obtain that when the mean binding time is significantly longer than the
mean mobile time, transient anomalous diffusion is observed at short and
intermediate time scales, with a strong dependence on the fraction of initially
mobile and immobile particles. We unveil a Laplace distribution of particle
displacements at relevant intermediate time scales. For any initial fraction
of mobile particles, the respective mean squared displacement displays a
plateau. Moreover, we demonstrate a short-time cubic time dependence of the
mean squared displacement for immobile tracers when initially all particles
are immobile.
\end{abstract}

\section{Introduction}

Already in the 1960ies there was considerable interest in the transport of
chemical tracers, especially pesticides, nitrates, or heavy metals through
water-carrying layers of the soil \cite{biggar}. A typical description for
such contaminant transport was the diffusion-advection equation (sometimes
called convective-dispersive equation) \cite{lapidus}
\begin{equation}
\label{dae}
\frac{\partial}{\partial t}C(x,t)=D\frac{\partial^2}{\partial x^2}C(x,t)-
v\frac{\partial}{\partial x}C(x,t),
\end{equation}
where $C(x,t)$ is the contaminant concentration at distance $x$ after time
$t$, $v$ is an advection velocity chosen as zero in the following, and $D$
the diffusion constant (dispersion coefficient typically measured in units
of $\mathrm{cm}^2/\mathrm{day}$). Measurements revealed, however, that not
all of the contaminant concentration was mobile at any given time, but that
a fraction could be (transiently) trapped in stagnant volumes. Building on
earlier models by Deans \cite{deans} and Coats and Smith \cite{coats}, van
Genuchten and Wierenga analyse the exchange between a mobile ($C_{\mathrm{
m}}(x,t)$) and immobile ($C_{\mathrm{im}}(x,t)$) fraction \cite{genuchten}.
In many geophysical systems equations of the type (\ref{dae}) are modified
to account for anomalous transport, in which molecular transport no longer
follows the linear time dependence $\langle\Delta x^2(t)\rangle=\langle x^2
(t)\rangle-\langle x(t)\rangle^2=2Dt$ of Brownian motion, but follows laws
of the type $\langle\Delta x^2(t)\rangle=2D_{\alpha}t^{\alpha}$, for which
$\alpha\neq1$ \cite{report}. Indeed such transport anomalies were found on
large field experiments, up to kilometre scales \cite{made,brian}. In such
systems the mobile-immobile transport model is replaced by models in which
generalised transport terms are incorporated \cite{schumer,doerries}. This
type of models,  in contrast to equation (\ref{dae}),  is characterised by
non-Gaussian distributions \cite{report}.

Motivated by concrete biological examples we here study a seemingly simple
version of the mobile-immobile transport model,  in which particles switch
between a freely diffusive phase and an immobile, stagnant phase. Even for
the Poissonian switching dynamics considered here between the mobile and
immobile phases and for biologically relevant parameters, we demonstrate
the existence of a significant, transient anomalous-diffusive regime with
 distinct non-Gaussian displacement distribution.

In fact, various components of biological cells, including tau proteins,
synaptic vesicles in hippocampal neurons, glucocorticord receptors, calcium
sensing proteins and transcription factors at the junction of the endoplasmic
reticulum and the plasma membrane undergo diffusion with intermittent
immobilisation \cite{igaev,janning,yeung,sprague,liu,mazza,chen,wu}. We here
focus on tau proteins, that intermittently bind to microtubules in axons of
neuronal cells and are then immobilised, as schematically depicted in figure
\ref{figscheme}. Tau proteins stabilise microtubules that give structure to
cells \cite{kolarova}. Alzheimer's disease is associated with tau proteins
losing the ability to bind to microtubules \cite{kolarova,guo2017roles}. This
effectively destabilises the microtubules and leads to neurodegeneration
\cite{kolarova,guo2017roles}. Due to the extremely elongated shape of the axon
the motion of tau proteins can be effectively described in one dimension
\cite{igaev}. If the immobilisation time follows an exponential distribution
with mean $\tau_\mathrm{im}$ and tracers immobilise with rate $\tau_\mathrm{
m}^{-1}$, i.e., a Poissonian dynamics, as assumed in \cite{igaev}, the motion
can be described by the mobile-immobile model
\begin{eqnarray}
\frac{\partial}{\partial t}n_\mathrm{m}(x,t)&=&-\frac{1}{\tau_\mathrm{m}}
n_\mathrm{m}(x,t)+\frac{1}{\tau_\mathrm{im}}n_\mathrm{im}(x,t)+D\frac{
\partial^2}{\partial x^2}n_\mathrm{m}(x,t)\nonumber\\
\frac{\partial}{\partial t}n_\mathrm{im}(x,t)&=&-\frac{1}{\tau_\mathrm{im
}}n_\mathrm{im}(x,t)+\frac{1}{\tau_\mathrm{m}}n_\mathrm{m}(x,t).
\label{eq00}
\end{eqnarray}
Here $n_\mathrm{m}(x,t)$ and $n_\mathrm{im}(x,t)$ denote the line densities of
mobile and bound tau proteins, respectively, with physical dimension $[1/
\mathrm{length}]$. The diffusion coefficient of the
mobile tracers is $D$. Since we are dealing with a system of non-interacting
particles, we use a probabilistic formulation according to which the total
concentration $n_\mathrm{tot}(x,t)=n_\mathrm{m}(x,t)+ n_\mathrm{im}(x,t)$ is
normalised to unity, $\int_{-\infty}^{\infty}n_\mathrm{tot}(x,t)dx=1$. The
line densities $n_\mathrm{m}(x,t)$ and $n_ \mathrm{im}(x,t)$ are then
the respective fractions. Equations (\ref{eq00}) were analysed in three
dimensions for an equilibrium fraction of initially mobile tracers, finding
Fickian yet non-Gaussian diffusion \cite{mora2018brownian}. Accordingly, the
mean squared displacement (MSD) of the total concentration $n_\mathrm{tot}$
grows linearly at all times, and under certain conditions a non-Gaussian
distribution emerges \cite{mora2018brownian}.

Such Fickian yet non-Gaussian diffusion has been shown to occur for the
motion of colloidal beads on phospholipid bilayer tubes, molecules at
surfaces, and colloids in a dense matrix of micropillars, where the
colloids can get trapped in pockets \cite{wang,skaug,chakraborty}.
Fickian-yet non-Gaussian diffusion with a finite correlation time beyond
which the displacement probability density function (PDF) crosses over to
a Gaussian with an effective diffusivity, arises in diffusing-diffusivity
models, in which the diffusivity of individual tracers varies stochastically
over time \cite{chechkin,chubynsky,wang1,sposini,lanoiselee,jain}. Direct
examples for such randomly evolving diffusion coefficients (mobilities) are
indeed known from lipids in protein-crowded bilayer membranes \cite{ilpo},
shape-shifting protein molecules \cite{eiji}, or (de)polymerising oligomer
chains \cite{fulvio,mario}. In other systems an intermittent plateau emerges
in the MSD, for instance, for two-dimensional fluids confined in a random
matrix of obstacles or a porous cavity, in which trapping in finite pockets
plays a key role \cite{skinner,wang2,slezak}. We also mention plateaus in
the MSD of both two- and three-dimensional isotropic Lennard-Jones binary
liquids \cite{sandalo}. In most of the systems mentioned
here the PDF crosses over from an exponential (Laplace) PDF to a Gaussian. In
the following we explicitly show how a Laplace distribution with fixed scale
parameters arises at intermediate time scales in our mobile-immobile model,
paired with transient anomalous diffusion.

In what follows we consider three initial conditions, an equilibrium fraction
of mobile tracers and a scenario in which initially all tracers are mobile
or immobile. These experimentally feasible situations significantly change
the diffusion at short and intermediate time scales, at which apparent
anomalous diffusion arises with slow-down and plateau-like behaviour, or
ballistic diffusion, respectively. Together with the transient non-Gaussian
displacement PDF this behaviour is remarkably rich, given the simplicity of
the governing equation (\ref{eq00}). We individually analyse the motion of the
mobile and immobile population of tracers, made possible by the formulation
of separate densities for mobile and immobile particles in this modelling
approach. One physical incentive to do so is that the function of the tau
proteins depends on their binding state \cite{kolarova}. Only bound tau
proteins stabilise microtubules, or transcription factors modulate gene
expression when bound to the DNA \cite{kolarova,liu}. In some situations
only the mobile or immobile tracers can be measured. An example is given
by combining total internal reflection fluorescence microscopy with
fluorescently labelled single stranded DNA, that binds to the microscope
cover slip \cite{peterson2016single}.

We present general results for the mobile and immobile concentrations and
the MSD for arbitrary fractions of initially mobile tracers in section
\ref{model}. Sections \ref{chmobile}, \ref{chimmo}, and \ref{cheq} present
concrete results and detailed discussions for different fractions of initial
mobile particle concentrations; respectively, we start with the cases when all
tracers are initially mobile and immobile and commence with an equilibrium
fraction of mobile tracers. We conclude in section \ref{concl}.

\section{Model and general solutions}
\label{model}

\begin{figure}
\centering
\includegraphics[width=.8\textwidth]{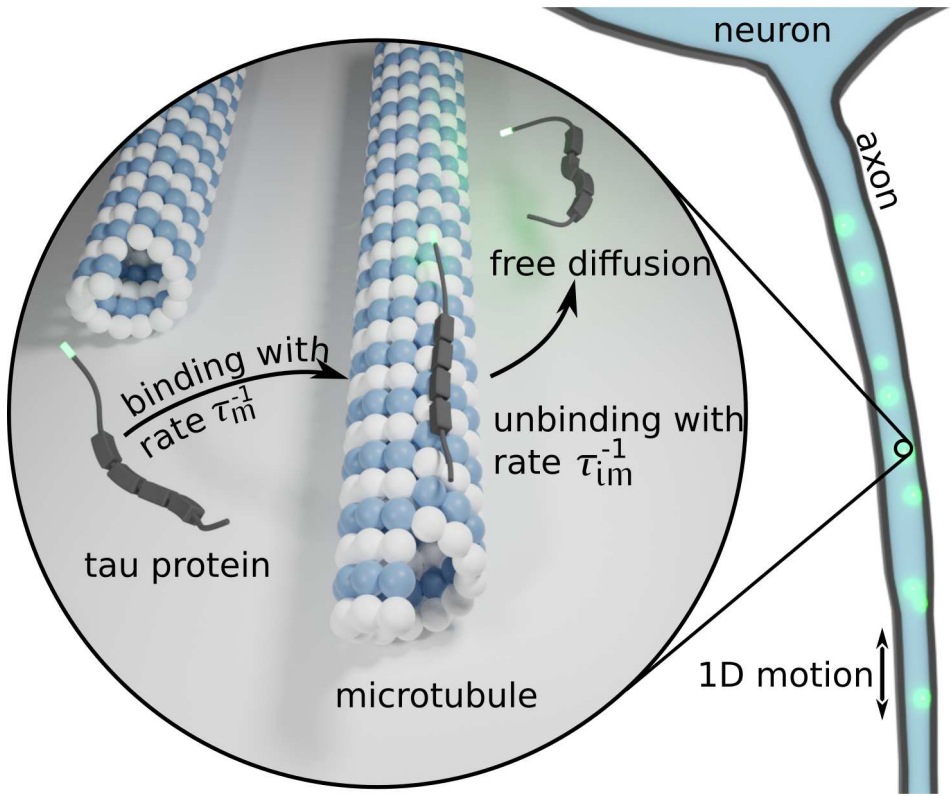}
\caption{Schematic of tau protein dynamics in axons of neuronal cells. Diffusing
tau proteins bind to longitudinally aligned  microtubules inside the axon with
the rate $\tau_\mathrm{m}^{-1}$. Upon
binding they remain immobile for the average duration $\tau_\mathrm{im}$ and
unbind with the rate $\tau_\mathrm{im}^{-1}$. The green markers represent
fluorescent proteins attached to the tau proteins. Due to the elongated shape
of the axons the tau protein dynamic can effectively be described in one
dimension. In our model we assume a homogeneous binding site density.}
\label{figscheme}
\end{figure}

We consider the mobile-immobile model equations (\ref{eq00}) for the initial
conditions $n_\mathrm{m}(x,0)=f_\mathrm{m}\delta(x)$ and $n_\mathrm{im}(x,0)
=f_\mathrm{im}\delta(x)$, where $f_\mathrm{m}$ and $f_\mathrm{im}$ denote the
fractions of initially mobile and immobile tracers, respectively, with the
normalisation $f_\mathrm{m}+f_\mathrm{im}=1$. This formulation is suitable
for typical single-particle tracking experiments used in biological and
soft matter systems. They are also relevant for geophysical experiments, in
which point-like injection of tracers are used. In this section we keep the
fractions $f_\mathrm{m}$ and $f_\mathrm{im}$ arbitrary and choose specific
values in the following three sections.

In what follows, we use the concrete
parameters $D=13.9$ $(\mu\mathrm{m})^2/\mathrm{sec}$, $\tau_\mathrm{m}=0.16
\mbox{ sec}$, and $\tau_\mathrm{im}=7.7\mbox{ sec}$ from \cite{igaev} in all
figures and neglect the vanishingly small advection velocity $v=0.002$ $\mu
\mathrm{m}/\mathrm{sec}$.\footnote{The slow directed motion only plays a
role when very long times are considered \cite{scholz,mercken,igaev}.}
Let us briefly address the experimental origin of the time scale
separation between $\tau_\mathrm{m}$ and $\tau_\mathrm{im}$.
From single particle tracking experiments of single-stranded DNA or tau
proteins, immobilisation times during the particle motion can be extracted
\cite{peterson2016single,janning}. The experiments for the tau proteins in
\cite{janning} provide two-dimensional information and revealed relatively
short residence times of the tau proteins on the microtubules, as compared
to mobile times \cite{janning}. In contrast, the fluorescence decay after
photoactivation (FDAP) experiment in one dimension along the axon direction,
here denoted as the $x$ variable, reveal long residence times and short mobile
periods: $\tau_\mathrm{im}\approx48\tau_\mathrm{m}$ \cite{igaev}.  This seeming
contradiction can be resolved when examining more closely the two-dimensional
trajectories in the supplementary material of \cite{janning}. Namely, the
microtubules inside the axon are aligned in parallel to the axon axis, as also
shown in our schematic \ref{figscheme}. While a single binding event is short,
an unbound particle quickly rebinds to a parallel, close-by microtubule
after a short distance covered by diffusion perpendicular to the axon axis.
This perpendicular motion does not contribute to the one-dimensional motion in
$x$ direction and thus, while individual binding times are relatively short,
\emph{effective\/} binding times appear much longer in the projection to
one dimension. Since we are only interested in the one-dimensional motion
we use the parameters of \cite{igaev} and hence long immobilisation times.

\subsection{Mobile and immobile concentration profiles}
\label{chgeneraln}

We consider the Fourier-Laplace transform of the concentrations and solve
for $n_\mathrm{m}(k,s)$, $n_\mathrm{im}(k,s)$ and $n_\mathrm{tot}(k,s)$ in
expressions (\ref{eq01}) and (\ref{eq02}), in which the Fourier wave number $k$
corresponds to the distance $x$ in real space, and the Laplace variable $s$ is
conjugated to time $t$, see \ref{secag} for details. We denote functions in
Fourier or Laplace space solely by replacing the explicit dependencies on
the respective arguments. The relations in Fourier-Laplace domain can be
Fourier-inverted, and we obtain the expressions 
\begin{eqnarray}
	\label{eq51}
\fl n_\mathrm{m}(x,s)=&\left(f_\mathrm{m}+f_\mathrm{im}\frac{1}{1+s\tau_
\mathrm{im}}\right)\frac{1}{\sqrt{4\phi(s)D}}e^{-\sqrt{\phi(s)/D}|x|}\\
\label{eq52}
\fl n_\mathrm{im}(x,s)=&\left(f_\mathrm{m}+f_\mathrm{im}\frac{1}{1+s\tau_
\mathrm{im}}\right)\frac{\tau_\mathrm{im}/\tau_\mathrm{m}}{1+s\tau_\mathrm{
im}}\frac{1}{\sqrt{4\phi(s)D}}e^{-\sqrt{\phi(s)/D}|x|}+f_\mathrm{im}
\frac{\tau_\mathrm{im}}{1+s\tau_\mathrm{im}}\delta(x)\\
\fl n_\mathrm{tot}(x,s)=&\frac{f_\mathrm{m}+f_\mathrm{im}\frac{1}{1+s\tau_
\mathrm{im}}}{s}\phi(s)\frac{1}{\sqrt{4\phi(s)D}}e^{-\sqrt{4\phi(s)/D}|x|}
+f_\mathrm{im}\frac{\tau_\mathrm{im}}{1+s\tau_\mathrm{im}}\delta(x)
\label{eq04}
\end{eqnarray}
as functions of $x$ and $s$ with $\phi(s)=s[1+\tau_\mathrm{im}
\tau_\mathrm{m}^{-1}/(1+s\tau_\mathrm{im})]$. These expressions are valid
for all $s$ and hence for all times $t$. A numerical Laplace inversion
then provides the densities for any specified time. Remarkably, it turns out
that the density of mobile tracers, that were initially immobile, is
proportional to the density of immobile tracers, that were initially mobile.
This can be seen by setting $f_\mathrm{m}=0$ or $f_\mathrm{im}=0$ in
(\ref{eq51}) and (\ref{eq52}), respectively.  This proportionality holds for all
$s$ and hence at all times.
We obtain the long-time Gaussian limit of the full
concentration in \ref{chal},
\begin{equation}
n_\mathrm{tot}(x,t)\sim\frac{1}{\sqrt{4\pi D_\mathrm{eff}t}}\exp\left(
-\frac{x^2}{4 D_\mathrm{eff}t}\right),\quad t\gg\tau_\mathrm{m},\tau_
\mathrm{im},
\label{eq30}
\end{equation}
with $D_\mathrm{eff}=D/(1+\tau_\mathrm{im}/\tau_\mathrm{m})$. 
Note that for asymptotic equalities we use the $\sim$ symbol.
In fact, independent of the ratio $f_\mathrm{m}$ and $f_\mathrm{im}$ we
asymptotically obtain a Gaussian distribution in which the diffusivity is
reduced to the effective diffusivity $D_\mathrm{eff}$. The mobile and immobile
concentrations are asymptotically equivalent to (\ref{eq30}) up to a scalar
$f_j^\mathrm{eq}$ defined below \cite{doerries}.

\subsection{Moments}

In general, the fractions $\overline{n}_\mathrm{m}$ and 
$\overline{n}_\mathrm{im}$ of mobile and immobile tracers, initially fixed
as $f_\mathrm{m}$ and $f_\mathrm{im}$, change over time. To
obtain the respective numbers, we integrate the tracer densities over space.
This corresponds to setting $k=0$ in the Fourier-Laplace transforms $n_
\mathrm{m}(k,s)$ and $n_\mathrm{im}(k,s)$ of the densities. After Laplace
inversion we find
\begin{eqnarray}
\overline{n}_\mathrm{m}(t)&=&\frac{\tau_\mathrm{m}}{\tau_\mathrm{m}+\tau_{im}}
+\frac{f_\mathrm{m}\tau_\mathrm{im}-f_\mathrm{im}\tau_{m}}{\tau_\mathrm{m}
+\tau_\mathrm{im}}\exp\left(-[\tau_\mathrm{m}^{-1}+\tau_\mathrm{im}^{-1}]t
\right),\\
\overline{n}_\mathrm{im}(t)&=&\frac{\tau_\mathrm{im}}{\tau_\mathrm{m}+\tau_\mathrm{im}}
-\frac{f_\mathrm{m}\tau_\mathrm{im}-f_\mathrm{im}\tau_{m}}{\tau_\mathrm{m}+
\tau_\mathrm{im}}\exp\left(-[\tau_\mathrm{m}^{-1}+\tau_\mathrm{im}^{-1}]t
\right),
\label{eq40}
\end{eqnarray}
with $\overline{n}_\mathrm{m}(t)+\overline{n}_\mathrm{im}(t)=1$.  In the
long-time limit $t\gg \tau_\mathrm{m},\tau_\mathrm{im}$ the fractions of mobile
and immobile tracers reach the stationary values
$f_\mathrm{m}^\mathrm{eq}=\tau_\mathrm{ m}/(\tau_\mathrm{m}+\tau_\mathrm{im})$
and $f_\mathrm{im}^\mathrm{eq}
=\tau_\mathrm{im}/(\tau_\mathrm{m}+\tau_\mathrm{im})$, respectively.
Our approach of splitting
the total concentration into mobile and immobile fractions allows us to
calculate the moments of the unbound, bound, and total tau protein
distributions individually,
\begin{equation}
\langle x^2(t)\rangle_j=\frac{1}{\overline{n}_j(t)}\int_{-\infty}^\infty x^2n_j(x,t)dx,
\label{eq43}
\end{equation}
where $j$ stands for $\mathrm{m}$, $\mathrm{im}$, and $\mathrm{tot}$
\cite{doerries}. To shorten the notation, we
use $\langle x^2(t)\rangle=\langle x^2(t)\rangle_\mathrm{tot}$ in the
remainder of this work. Using the Laplace inversion of 
\begin{equation}
\left.\frac{\partial^2}{\partial k^2}n_\mathrm{tot}(k,s)\right|_{k=0}=
\langle x^2(s)\rangle,
\label{eq06}
\end{equation}
we obtain the expression
\begin{equation}
\langle x^2(t)\rangle= 2D_{\mathrm{eff}}t
+2D\tau_\mathrm{im}\frac{f_\mathrm{m}\tau_\mathrm{im}/\tau_\mathrm{m}-
f_\mathrm{im}}{(1+\tau_\mathrm{im}/\tau_\mathrm{m})^2}\left(1-e^{-(
\tau_\mathrm{m}^{-1}+\tau_\mathrm{im}^{-1})t}\right)
\label{eq03}
\end{equation}
for the second moment. In the next section we consider the initial conditions,
when all tracers are initially mobile. This is chosen for didactic purposes, as
this initial condition shows the plateau in the MSD and intermittent Laplace 
distribution most clearly. In section \ref{chimmo} we consider immobile initial
conditions and finally consider equilibrium initial conditions in section
\ref{cheq}, where the effects discussed in earlier sections are present at 
the same time.

\section{All tracers initially mobile}
\label{chmobile}

We now consider the initial condition when all tracers are mobile, i.e.,
$n_\mathrm{m}(x,0)=\delta(x)$ and $n_\mathrm{im}(x,0)=0$. This initial
condition does not correspond to the experiment carried out by \cite{igaev}.
However, this situation could be realised experimentally, e.g., by using
the method of injection of fluorescently labelled tau proteins
\cite{kreis1982microinjection}.  
In what follows we repeatedly use the time scale separation
$\tau_\mathrm{m}\ll\tau_\mathrm{im}$ observed for tau proteins and also
relevant to other systems.

\subsection{Concentration}

\begin{figure}
\centering
\includegraphics{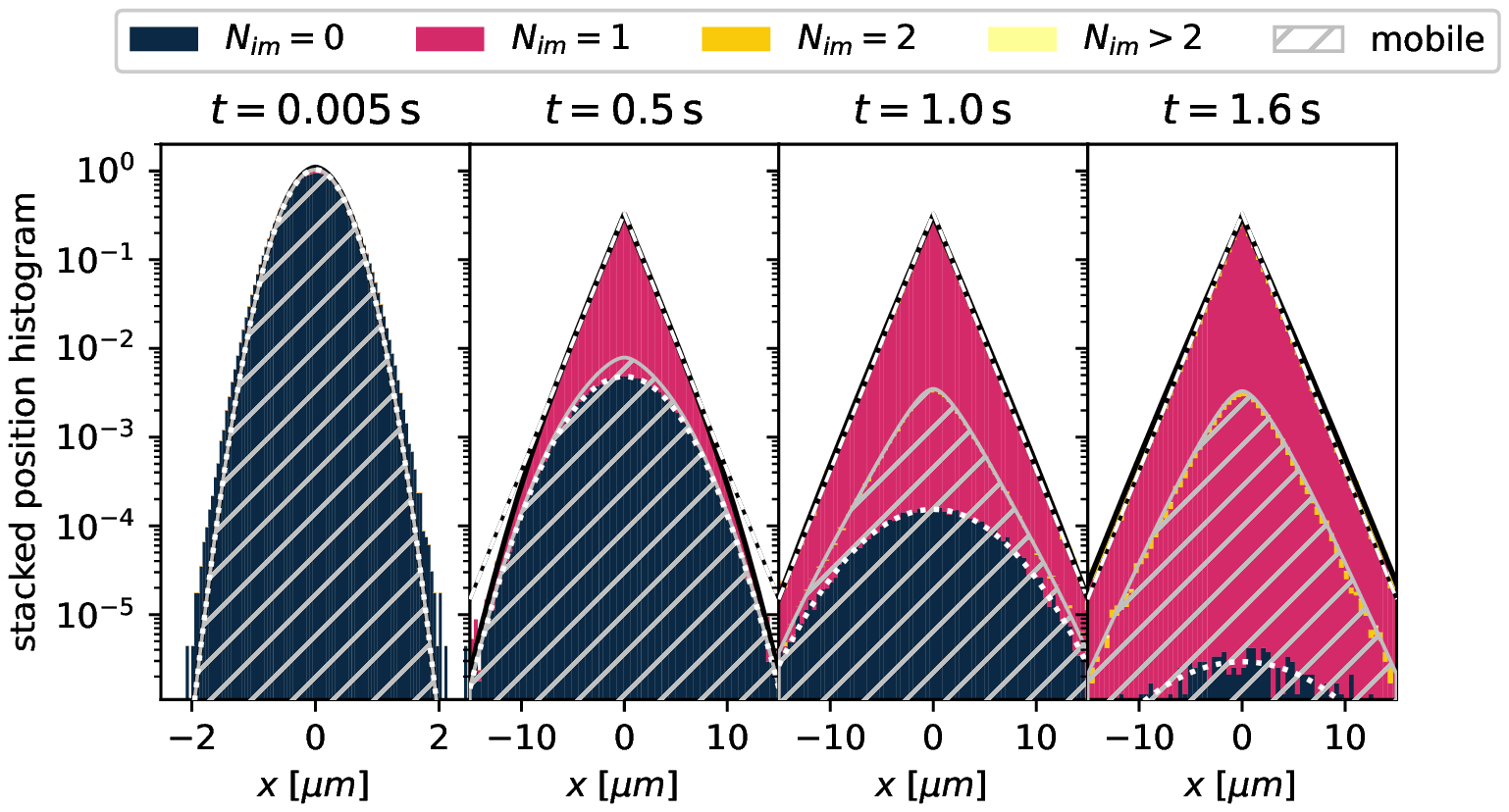}
\includegraphics{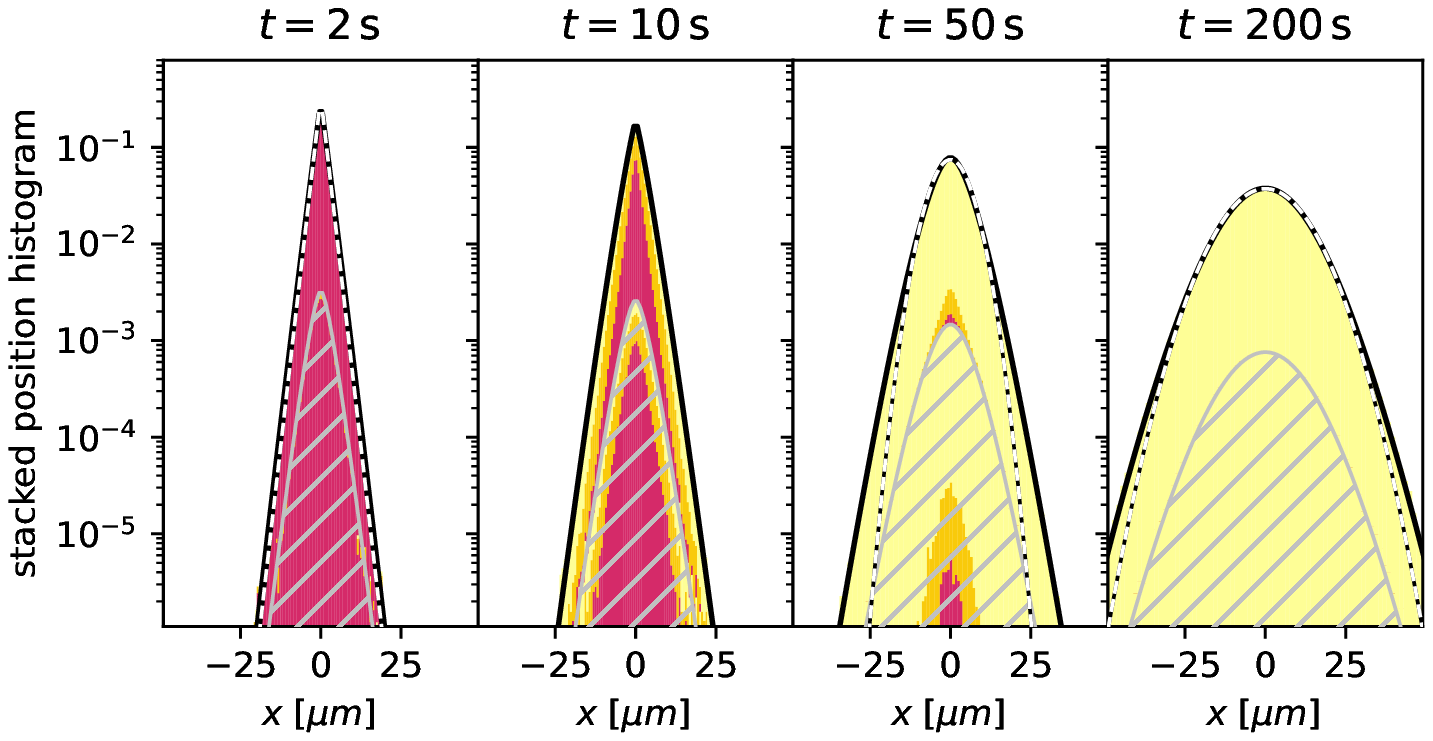}
\caption{Concentration profiles for mobile initial conditions. The solid black
line shows $n_\mathrm{tot}(x,t)$, the grey striped area $n_\mathrm{m}(x,t)$
obtained via Laplace inversion of relations (\ref{eq51}) and (\ref{eq04}).
Colours indicate the number of immobilisation events of particles from a
Brownian dynamics simulation with $5\times10^6$ trajectories in a stacked
histogram. The striped area denotes mobile particles and the white dotted
line denotes initially mobile tracers that have not yet been immobilised up to
the indicated time $t$ (\ref{eq25}); this result almost
coincides with the full concentration in the top left panel.  For
$t=0.5\,\mathrm{sec}$ to $2\,\mathrm{sec}$ the white dashed line shows the
Laplacian (\ref{eq39}), and for $t=50\,\mathrm{sec}$ and $200 \,\mathrm{sec}$ it
shows the long-time Gaussian (\ref{eq30}).}
\label{figProfile}
\end{figure}

We calculate the densities at short, intermediate and long times.
In \ref{chsl} we obtain the Gaussian 
\begin{equation}
\label{eq44}
n_\mathrm{tot}(x,t)\sim\frac{1}{\sqrt{4\pi Dt}}\exp\left(-\frac{x^2}{4Dt}
\right)
\end{equation}
in the short time limit $t\ll\tau_\mathrm{m},\tau_\mathrm{im}$.
The Gaussian (\ref{eq44}) can be seen in
figure \ref{figProfile} in the top left panel. 
In this figure $n_\mathrm{m}(x,t)$, $n_\mathrm{tot}
(x,t)$, and a histogram are shown. The densities are obtained from Laplace 
inversions of the expressions in Laplace space (\ref{eq04}), while the
histogram is obtained from simulations, 
and colours denote the number of immobilisation events $N_\mathrm{im}$.
Initially, all particles are mobile and diffuse freely, as denoted by the black
colouring.

The concentration of freely diffusing
particles that have not immobilised yet, i.e., have zero immobilisation
events $N_\mathrm{im}=0$, is given by the PDF of free Brownian motion
multiplied by the probability of not having immobilised, i.e.,
\begin{equation}
n_\mathrm{m}(x,t|N_\mathrm{im}=0)=\frac{\exp(-t/\tau_\mathrm{m})}{\sqrt{4\pi
Dt}}\exp\left(-\frac{x^2}{4Dt}\right).
\label{eq25}
\end{equation}
These mobile tracers immobilise with the position dependent rate $n_\mathrm{
m}(x,t|N_\mathrm{im}=0)/\tau_\mathrm{m}$. Integrating from $t'=0$ to $t'=t$,
we obtain in the limit $t\ll\tau_\mathrm{im}$ (i.e., at short and intermediate
times) that
\begin{eqnarray}
\nonumber
\fl n_\mathrm{im}(x,t\ll\tau_{\mathrm{im}})&\sim&\int_0^t
\frac{\exp(-t'/\tau_\mathrm{m})/\tau_\mathrm{m}}{\sqrt{4\pi Dt'}}
\exp\left(-\frac{x^2}{4 Dt'}\right)dt'\\
\nonumber
&=&\frac{\exp\left(-|x|/\sqrt{D\tau_\mathrm{m}}\right)}{\sqrt{4D\tau_
\mathrm{m}}}\frac{1-\mathrm{erf}\left(|x|/\sqrt{4Dt}-\sqrt{t/\tau_\mathrm{
m}}\right)}{2}\\
&&-\frac{\exp(|x|/\sqrt{D\tau_\mathrm{m}})}{\sqrt{4D\tau_\mathrm{m}}}
\frac{1-\mathrm{erf}\left(|x|/\sqrt{4Dt}+\sqrt{t/\tau_\mathrm{m}}\right)}{2}.
\label{eqer}
\end{eqnarray}
Comparing (\ref{eqer}) with the Laplace inversion of $n_\mathrm{im}(x,s)$
(\ref{eq04}) in figure \ref{figcimer} we find very good agreement in the
relevant range $t\ll\tau_\mathrm{im}$ \footnote{Equations (\ref{eq25}) and 
(\ref{eqer}) can also be obtained by taking the limit 
$\tau_\mathrm{im}\to\infty$ in (\ref{eq00}) and solving the equations
directly.}.  For the total density we obtain by adding $n_\mathrm{m}(x,t)$
(\ref{eq44}) and $n_\mathrm{im}(x,t)$ (\ref{eqer}) 
\begin{eqnarray}
n_\mathrm{tot}(x,t)&\sim&\frac{\exp(-t/\tau_\mathrm{m})}{
\sqrt{4\pi Dt}}\exp\left(-\frac{x^2}{4Dt}\right)
+n_\mathrm{im}(x,t\ll\tau_\mathrm{im}),\quad t\ll \tau_\mathrm{im}
\label{eq50}
\end{eqnarray}
for the full tracer density. For $t\ll \tau_\mathrm{m}$ we recover the Gaussian
(\ref{eq44}) from (\ref{eq50}), while for $\tau_\mathrm{m}\ll t
\ll\tau_\mathrm{im}$ the distribution is distinctly non-Gaussian, as shown
in figure \ref{figProfile}.    Up to around $t=0.6\,\mathrm{sec}$,
the motion of the free tracers is dominated by the Gaussian
$n_\mathrm{m}(x,t|N_\mathrm{im}=0)$, see
(\ref{eq25}), which spreads like free Brownian particles, shown as a white
dotted line in figure \ref{figProfile}.  At around $t=1.6\,\mathrm{ sec}$, most
of the tracers with $N_\mathrm{im}=0$ immobilised and the majority of mobile
tracers were immobile exactly once ($N_\mathrm{im}=1$) and transitioned back to
the mobile zone, as shown by the red area.  Due to the immobilisation, these
tracers have moved less than the free particles with $N_\mathrm{im}=0$ and a
Laplace distribution emerges in the centre.
For $x\ll t\sqrt{D/\tau_\mathrm{m}}$ and $t\gg\tau_\mathrm{m}$ we can use
the asymptotic $\lim_{x\to\infty}\mathrm{erf}(x)=-\lim_{
x\to\infty}\mathrm{erf}(-x)=1$ in $n_\mathrm{im}(x,t\ll\tau_{\mathrm{im}})$,
equation
(\ref{eqer}), and obtain from $n_\mathrm{tot}(x,t)$ (\ref{eq50}) the expression
\begin{equation}
\label{eq39}
n_\mathrm{tot}(x,t)\sim\frac{1}{\sqrt{4D\tau_\mathrm{m}}}\exp
\left(-\frac{|x|}{\sqrt{D\tau_\mathrm{m}}}\right),
\end{equation}
in the intermediate time regime $\tau_\mathrm{m}\ll t\ll\tau_\mathrm{im}$.
Combining the conditions $t\ll\tau_\mathrm{m}$ and $x\ll t\sqrt{D/\tau_
\mathrm{im}}$ leads to $x\ll\tau_\mathrm{im}\sqrt{D/\tau_\mathrm{m}}=71\,
\mu\mathrm{m}$, which is large compared to the standard deviation
$\sqrt{2D\tau_\mathrm{m}}=2.1\,\mu\mathrm{m}$ of the Laplace distribution
(\ref{eq39}). This means that the distribution follows such a Laplace
shape for a large range of positions. The total concentration, in turn,
therefore follows a Laplace distribution with fixed
parameters. This is a pronounced deviation from a Gaussian distribution. This
result can also be obtained from calculations in Laplace space, as shown in
\ref{chagi}. In contrast, for times significantly longer than $\tau_\mathrm{
im}$, many immobilisations take place, as shown by the bright yellow area in
figure \ref{figProfile}, where the distribution follows the Gaussian
(\ref{eq30}) with the effective diffusivity $D_\mathrm{eff}=D/(1+\tau_
\mathrm{im}/\tau_\mathrm{m})$.

\subsection{Mean squared displacement}

\begin{figure}
\centering
\includegraphics[width=\textwidth]{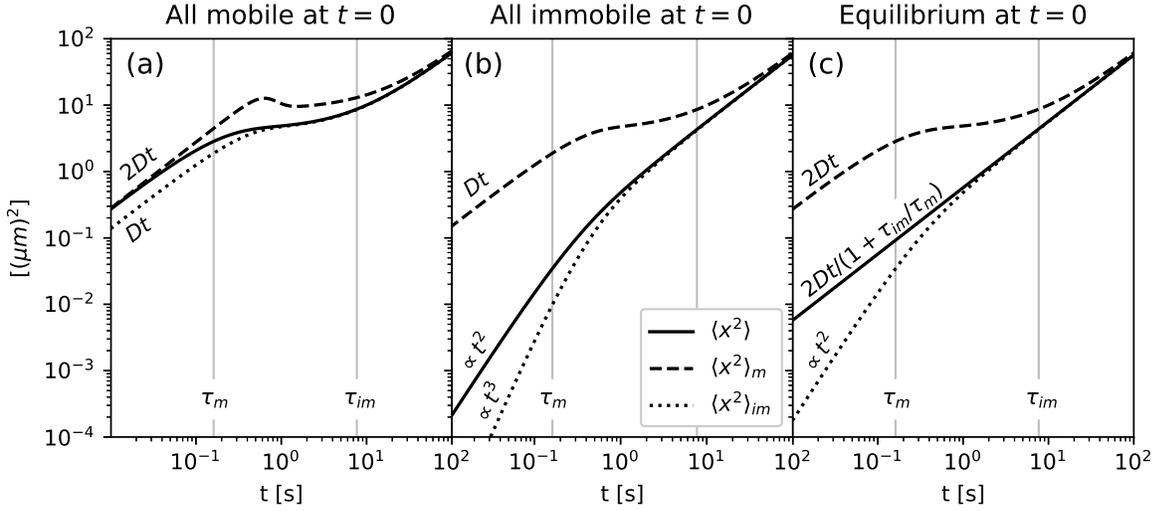} 
\caption{Second moments for different initial conditions on a log-log scale.
In panel (a) all tracers are initially mobile, as in section \ref{chmobile}.
After a linear growth, the second moment $\langle x^2(t)\rangle$ of all
tracers, equation (\ref{eq07}), shows a plateau for $\tau_\mathrm{m}\ll
t\ll\tau_\mathrm{im}$. The second moment of the mobile particles, equation
(\ref{eq23}), in panel (a) has a peak immediately before the total particle
moment and the mobile particle moment reach a plateau value. Immobile tracers
spread $\sim Dt$ at short times, and the second moment (\ref{eq24}) has a
plateau at intermediate times. In panel (b), all tracers are initially
immobile, as in section \ref{chimmo}. The second moment of all tracers,
equation (\ref{eq09}), grows $\sim Dt^2/\tau_\mathrm{im}$ at short times, due to
the decaying number of particles located at $x=0$. The immobile tracers spread
$\sim Dt^3/(3\tau_\mathrm{m}\tau_\mathrm{im})$ at short times, while the full
expression is given in equation (\ref{eq09}). The mobile tracers in (b)
spread exactly like the immobile tracers in (a), where all tracers are initially
mobile. Panel (c) shows the equilibrium case, section \ref{cheq}, in which the
second moment grows like $2Dt/(1+\tau_\mathrm{im}/\tau_\mathrm{m})$, equation
(\ref{eq05}) for all times. The mobile and immobile moments exactly match the
moments of the total distribution with mobile and immobile initial conditions,
respectively.}
\label{figxx}
\end{figure}

From the general expression for the MSD (\ref{eq03}) for immobile initial
conditions, we obtain  the expression
\begin{equation}
\langle x^2(t)\rangle=\frac{2D}{1+\tau_\mathrm{im}/\tau_\mathrm{
m}}\left[t+\frac{\tau_\mathrm{im}^2/\tau_\mathrm{m}}{1+\tau_\mathrm{im}/
\tau_\mathrm{m}}\left(1-e^{-(\tau_\mathrm{m}^{-1}+\tau_\mathrm{im}^{-1})t}
\right)\right].
\label{eq07}
\end{equation}
At intermediate times the MSD, expression (\ref{eq07}), exhibits a
plateau-like behaviour with the constant MSD
\begin{equation}
\langle x^2(t)\rangle\sim2D\tau_\mathrm{m},\quad\tau_\mathrm{m}\ll
t\ll\tau_\mathrm{im},
\end{equation}
corresponding to free Brownian particles that moved for the duration $\tau_
\mathrm{m}$. This requires the condition $\tau_\mathrm{m}\ll\tau_\mathrm{im}$,
which is satisfied in the tau protein case \cite{igaev}, with $\tau_\mathrm{m}
=0.16\,\mathrm{sec}$ and $\tau_\mathrm{im}=7.7\,\mathrm{sec}$. Such plateaus
are often found when tracers diffuse in porous media or for dynamics in
crowded membranes or environments with obstacles, in which the tracer can be
transiently confined \cite{wang1,skinner,slezak,ghosh,matti}. The MSD
(\ref{eq07}) is shown in figure \ref{figxx}(a) as the black solid line.

When calculating the moments
of the mobile and immobile tracers (\ref{eq43}), the time-dependent
normalisations of the tracer densities (\ref{eq40}),
\begin{eqnarray}
\overline{n}_\mathrm{m}(t)&=\frac{\tau_\mathrm{m}}{\tau_\mathrm{m}+\tau_\mathrm{im}}
\left[1+\tau_\mathrm{im}/\tau_\mathrm{m} e^{-(\tau_\mathrm{m}^{-1}+\tau_
\mathrm{im}^{-1})t)} \right]\\
\overline{n}_\mathrm{im}(t)&=\frac{\tau_\mathrm{im}}{\tau_\mathrm{m}+\tau_\mathrm{im}}
\left[ 1- e^{-(\tau_\mathrm{m}^{-1}+\tau_\mathrm{im}^{-1})t)} \right],
\label{eq18}
\end{eqnarray}
need to be taken into account, yielding the moments of the
mobile and immobile densities (\ref{eq43}) \cite{doerries}
\begin{eqnarray}
\langle x^2(t)\rangle_\mathrm{m}&=&\frac{2D}{(1+\tau_\mathrm{im}/\tau_
\mathrm{m})(1+\tau_\mathrm{im}/\tau_\mathrm{m}e^{-(\tau_\mathrm{m}^{-1}
+\tau_\mathrm{im}^{-1})t})}\Big[t\left(1+\frac{\tau_\mathrm{im}^2}{
\tau_\mathrm{m}^{2}}e^{-(\tau_\mathrm{m}^{-1}+\tau_\mathrm{im}^{-1})t}
\right)\nonumber\\
&&+\frac{2\tau_\mathrm{im}^2/\tau_\mathrm{m}}{1+\tau_\mathrm{im}/\tau_
\mathrm{m}}(1-e^{-(\tau_\mathrm{m}^{-1}+\tau_\mathrm{im}^{-1})t})\Big]
\label{eq23}
\end{eqnarray}
and 
\begin{eqnarray}
\langle x^2(t)\rangle_\mathrm{im}&=&\frac{2D}{1-e^{-(\tau_\mathrm{m}^{-1}
+\tau_\mathrm{im}^{-1})t}}\left[ \frac{t}{1+\tau_\mathrm{im}/\tau_\mathrm{
m}}\left(1-\frac{\tau_\mathrm{im}}{\tau_\mathrm{m}}e^{-(\tau_\mathrm{m}^{-1}
+\tau_\mathrm{im}^{-1})t}\right)\right.\nonumber\\
&&+\left.\frac{\tau_\mathrm{im}^2/\tau_\mathrm{m}-\tau_\mathrm{im}}{(1+
\tau_\mathrm{im}/\tau_\mathrm{m})^2}\left(1-e^{-(\tau_\mathrm{m}^{-1}+
\tau_\mathrm{im}^{-1})t}\right)\right].
\label{eq24}
\end{eqnarray}
As shown in figure \ref{figxx}, the mobile second moment exhibits a peak
at around $t=0.6\,\mathrm{sec}$, followed by a plateau. This peak arises
as the density of mobile tracers initially consists of mobile tracers that
have never immobilised. Once $t\gg\tau_\mathrm{m}$ the mobile density mainly
consists of tracers that were immobile (at least) once and mobilised, as
discussed above. Since the latter had less time to move, they have spread
less and the MSD temporarily decreases.

\begin{figure}
\centering
\includegraphics[width=\textwidth]{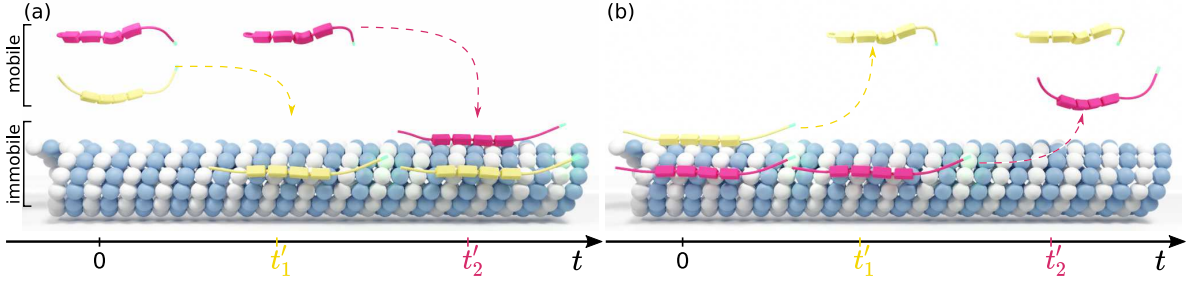}
\caption{Schematic showing the short-time behaviour of tracers for
mobile (a) and immobile initial conditions (b) at three snapshots of time.
In both panels, the tracers change the mobilisation state at times $t_1'$
and $t_2'$, respectively. For mobile initial conditions in (a), the number
of immobile tracers grows $\sim t/\tau_\mathrm{m}$ at short times. Namely,
the later a tracer immobilises, the longer it was previously mobile. In (b)
the number of mobile tracers grows $\sim t/\tau_\mathrm{im}$. Namely, the
earlier a tracer mobilises in (b), the longer it is mobile.}
\label{figImScheme}
\end{figure}

The immobile MSD (\ref{eq24}) has the short-time behaviour $\langle x^2(t)
\rangle_\mathrm{im}\sim Dt$ for $t\ll\tau_\mathrm{m},\tau_\mathrm{im}$. The
factor $\frac{1}{2}$ as compared to the mobile tracers arises because
immobile tracers effectively average over the history of the mobile tracers.
Namely, for $t'\ll\tau_\mathrm{m},\tau_\mathrm{im}$, mobile particles
immobilise with the constant rate $p(t')=1/\tau_\mathrm{m}$.
A particle that immobilised at time $t'$ before moved for the duration $t'$
and thus contributes $2Dt'$ to the second moment for $t>t'$, see figure
\ref{figImScheme}(a) for a schematic drawing. When averaging over different
mobile periods $t'$ and normalising with the fraction of immobile
tracers $\int_0^t p(t')dt'$, we obtain
\begin{equation}
\label{eq38}
\langle x^2(t)\rangle_\mathrm{im}\sim
2D\frac{\int_0^t t'p(t')dt'}{\int_0^t p(t')dt'}=
\frac{2D\int_0^tt'/\tau_\mathrm{m}
dt'}{t/\tau_\mathrm{m}}=Dt,\mbox{ for }
t\ll\tau_\mathrm{m},\tau_\mathrm{im}.
\end{equation}
As mentioned above,
the long-time limits of the MSDs of all densities remain equal to $2D_
\mathrm{eff}t$, regardless of the fractions $f_\mathrm{m}$ and $f_\mathrm{im}$.

\section{All tracers initially immobile}
\label{chimmo}

\begin{figure}
\centering
\includegraphics{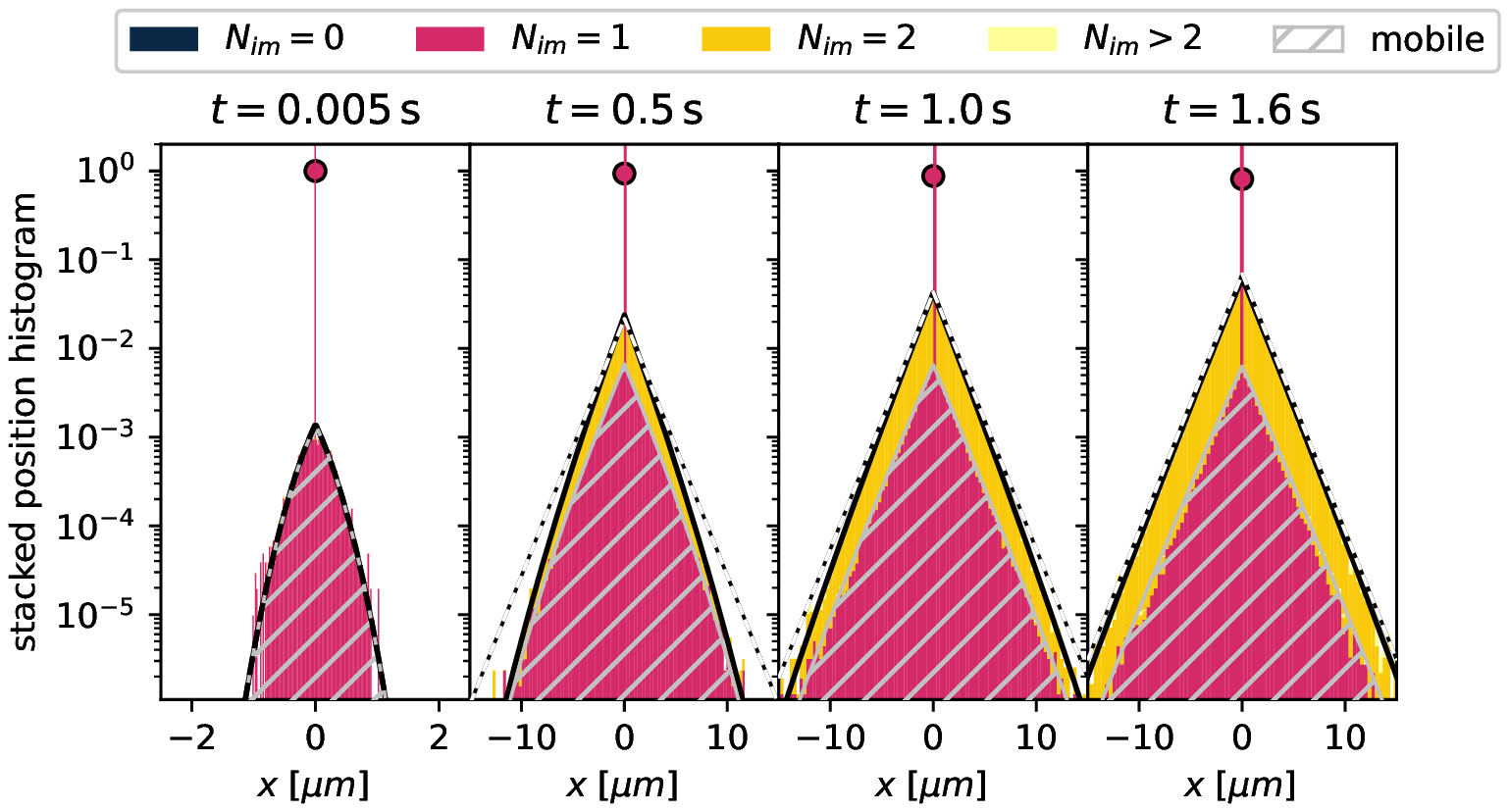}
\includegraphics{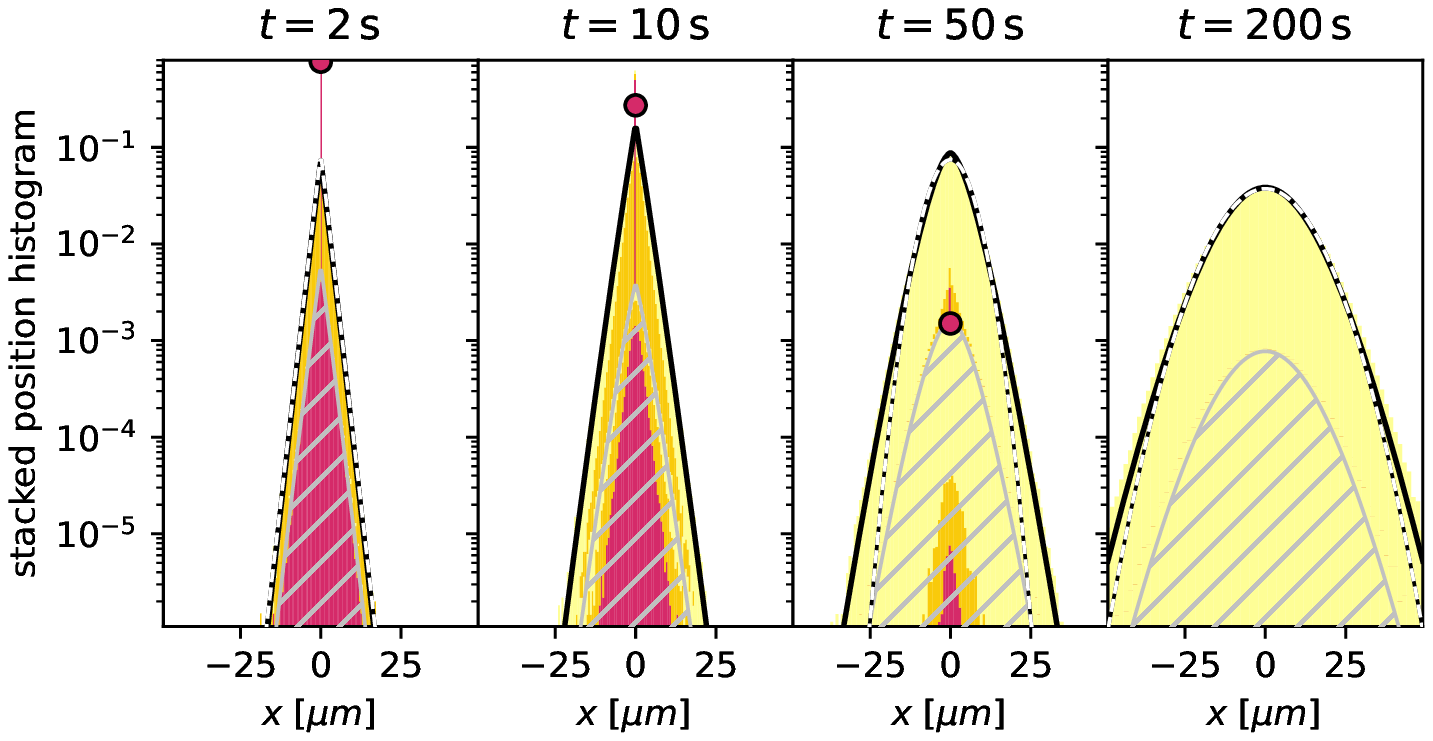}
\caption{Concentration profiles for immobile initial conditions, for a
description of the legend see figure \ref{figProfile}. The main difference to
the case of mobile initial conditions poses the peak of immobile tracers at
$x=0$ that have not moved up to time $t$, as shown by the circle. In addition
we here find a
pronounced relative increase of mobile particle numbers and the very slow spread
of immobile tracers at short times.  The short time approximation (\ref{eq45})
is shown as the black dashed line in the top left panel. For $t=0.5\,\mathrm{sec}$
to $10\,\mathrm{sec}$ the white dashed line shows the Laplacian (\ref{eq48})
with growing weight, and for $t=50\,\mathrm{sec}$ and $200 \,\mathrm{sec}$ it
shows the long-time Gaussian (\ref{eq30}).}
\label{fighistIm}
\end{figure}

We now discuss the case when all tracers are immobile at $t=0$,
$n_\mathrm{im}(x,0)=\delta(x)$ and $n_\mathrm{m}(x,0)=0$.

\subsection{Concentration}

In \ref{chsl}, we obtain the short-time behaviour
\begin{equation}
\fl n_\mathrm{tot}(x,t)\sim \frac{2t/\tau_\mathrm{im}}{\sqrt{4 \pi D t}}
e^{-\frac{x^2}{4Dt}}
-\frac{|x|(1-\mathrm{erf}\left(\frac{|x|}{\sqrt{4Dt}}\right)}
{2D\tau_\mathrm{im}}
+\left(1-\frac{t}{\tau_\mathrm{im}}\right)\delta(x),\mbox{ for }
t\ll\tau_\mathrm{m},\tau_\mathrm{im}
\label{eq45}
\end{equation}
by applying approximations for large $s$ in Laplace space.
Expression (\ref{eq45}) is shown in the left panel of figure \ref{fighistIm} 
as the black dashed line.
In particular note the distinctively non-Gaussian shape of the distribution in
contrast to the case of mobile initial conditions. The Gaussian in equation 
(\ref{eq45}) has the normalisation $\sim 2 t/\tau_\mathrm{im}$, while the 
second term has the normalisation $\sim -t/\tau_\mathrm{im}$, and thus the
whole expression (\ref{eq45}) is normalised to unity. In \ref{chagi}
we obtain the total density at intermediate times $\tau_\mathrm{m}\ll
t\ll\tau_\mathrm{im}$ 
\begin{equation}
\label{eq48}
n_\mathrm{tot}(x,t)\sim\frac{t/\tau_\mathrm{im}}{\sqrt{4D\tau_
\mathrm{m}}}\exp\left(-\frac{|x|}{\sqrt{D\tau_\mathrm{m}}}\right)
+\left(1-\frac{t}{\tau_\mathrm{im}}\right)\delta(x),
\end{equation}
as shown in figure \ref{fighistIm} in the top row (except for the leftmost
panel) as a black-white striped line. Compared to the mobile initial condition,
the coefficient of the Laplace distribution has the linear growth
$t/\tau_\mathrm{im}$.  Most tracers remain immobile at the origin at
$t=1.6\,\mathrm{sec}$.  In \ref{chImShortInt} we find expression  (\ref{eq006})
for $n_\mathrm{tot}(x,t)$, that is valid for $t\ll\tau_\mathrm{im}$ and contains
equations (\ref{eq45}) and (\ref{eq48}) as limits.  In figure \ref{fighistIm}
the lower panels show the transition from the Laplace distribution to the
Gaussian (\ref{eq30}).

\subsection{Mean squared displacement}

From the general expression for the MSD (\ref{eq03}), we obtain  the expression
\begin{eqnarray}
\langle x^2(t)\rangle=\frac{2D}{1+\tau_\mathrm{im}/\tau_\mathrm{
m}}\left[t-\frac{\tau_\mathrm{im}}{1+\tau_\mathrm{im}/\tau_\mathrm{m}}\left(1-
e^{-(\tau_\mathrm{m}^{-1}+\tau_\mathrm{im}^{-1})t}\right)\right].
\label{eq09}
\end{eqnarray}
The MSD (\ref{eq09}) has the ballistic short-time behaviour  
\begin{equation}
\label{eq10}
\langle x^2(t)\rangle\sim\frac{Dt^2}{\tau_\mathrm{im}}+O(t^3),
\quad t\ll\tau_\mathrm{m},\tau_\mathrm{im}.
\end{equation}
The Landau symbol $O(\cdot)$ represents higher order terms.
The ballistic behaviour at short times $t\ll\tau_\mathrm{im}$ arises because
the fraction $\exp(-t/\tau_\mathrm{im})\sim 1-t/\tau_\mathrm{im}$ of tracers
are immobile at $x=0$ and hence do not contribute to the second moment. For
$t'\ll\tau_\mathrm{m}\ll\tau_\mathrm{im}$, immobile particles mobilise with
the constant rate $p(t')=1/\tau_\mathrm{im}$.
A particle that mobilised at time $t'$ moved for the duration $t-t'$
and thus contributes $2D(t-t')$ to the second moment for $t>t'$, see figure
\ref{figImScheme}(a) for a schematic drawing. When integrating over different
mobilisation times $t'$ we find
\begin{equation}
\label{eq29}
\fl \langle x^2(t)\rangle\sim 2D\int_0^t (t-t')p(t')dt'
=2D\int_0^t\frac{t-t'}{\tau_\mathrm{im}}dt'=D\frac{t^2}{\tau_\mathrm{im}},
\quad t\ll\tau_\mathrm{m}\ll\tau_\mathrm{im}.
\end{equation}
We obtain the number of free
and bound tracers from the general expression (\ref{eq40}),
\begin{eqnarray}
\overline{n}_\mathrm{m}(t)&=\frac{\tau_\mathrm{m}}{\tau_\mathrm{m}+\tau_\mathrm{im}}
\left[1-e^{-(\tau_\mathrm{m}^{-1}+\tau_\mathrm{im}^{-1})t)}\right]\\
\overline{n}_\mathrm{im}(t)&=\frac{\tau_\mathrm{im}}{\tau_\mathrm{m}+\tau_\mathrm{im}}
\left[1+\tau_\mathrm{m}/\tau_\mathrm{im}e^{-(\tau_\mathrm{m}^{-1}+\tau_
\mathrm{im}^{-1})t)}\right].
\label{eq32}
\end{eqnarray}
This produces the normalisation of the immobile moment, and we find
\begin{equation}
\label{eq34}
\fl\langle x^2(t)\rangle_\mathrm{im}=\frac{2Dt}{1+\tau_\mathrm{m}/\tau_\mathrm{
im}}\frac{1+e^{-(\tau_\mathrm{m}^{-1}+\tau_\mathrm{im}^{-1})t}}{\tau_\mathrm{
im}/\tau_\mathrm{m}+e^{-(\tau_\mathrm{m}^{-1}+\tau_\mathrm{im}^{-1})t}}
-\frac{4D\tau_\mathrm{im}^2/\tau_\mathrm{m}}{(1+\tau_\mathrm{im}/\tau_\mathrm{
m})^2}\frac{1-e^{-(\tau_\mathrm{m}^{-1}+\tau_\mathrm{im}^{-1})t}}{\tau_\mathrm{
im}/\tau_\mathrm{m}+e^{-(\tau_\mathrm{m}^{-1}+\tau_\mathrm{im}^{-1})t}}.
\end{equation}
This MSD has the short-time behaviour $\langle x^2(t)\rangle_\mathrm{im}\sim
Dt^3/(3\tau_\mathrm{im}\tau_\mathrm{m})$ for $t\ll\tau_\mathrm{m},\tau_\mathrm{
im}$. The cubic scaling emerges as the only immobile tracers, that are not
located at the origin, have previously mobilised and then immobilised again.
The mobile concentration grows $\sim t/\tau_\mathrm{im}$ at short times $t\ll
\tau_\mathrm{im}$.
Integrating over the time $t'$ spent in the mobile phase yields the cubic
scaling
\begin{equation}
\label{eq35}
\fl\langle x^2(t)\rangle_\mathrm{im}\sim2D\int_0^t\frac{1}{\tau_\mathrm{m}} 
n_\mathrm{m}(t-t')t'dt'=\frac{2D}{\tau_\mathrm{m}\tau_\mathrm{im}}\int_0^t
(t-t')t'dt'=D\frac{t^3}{3\tau_\mathrm{m}\tau_\mathrm{im}},
\end{equation}
where in the first step we took the limit $t\ll\tau_\mathrm{m}$. 
Since the mobile concentration with immobile initial conditions is
proportional to the immobile concentration with mobile initial conditions,
$\langle x^2(t) \rangle_\mathrm{m}$ is equal to $\langle x^2(t)\rangle_
\mathrm{im}$ in (\ref{eq24}) with mobile initial conditions. This 
can be seen in figure \ref{figxx}(a) and \ref{figxx}(b). 
As for the mobile initial condition considered in section \ref{chmobile}, the
MSDs of all densities grow $\sim 2D_\mathrm{eff}t$ asymptotically.

\section{Equilibrium initial fractions of initial mobile tracers}
\label{cheq}

In this section we use the equilibrium values $n_\mathrm{m}(x,0)=f_\mathrm{m}
^\mathrm{eq}\delta(x)$ and $n_\mathrm{im}(x,0)=f_\mathrm{im}^\mathrm{eq}
\delta(x)$ as initial conditions.

\subsection{Concentration profiles} 

From the general expressions (\ref{eq51}) and (\ref{eq04}) for the densities
$n_\mathrm{m}(x,s)$ and $n_\mathrm{tot}(x,s)$ we find that the mobile
concentration of the equilibrium case discussed here is proportional
to the total concentration for the mobile initial condition in section
\ref{chmobile} at all times. To understand why this is true, we notice
that both concentrations at all times contain mobile tracers that were
initially mobile. Moreover, from equations (\ref{eq51}) and (\ref{eq04}) we
see that the mobile concentration of the equilibrium case contains initially
immobile tracers, while the total concentration contains immobile tracers,
that were initially mobile. In equations (\ref{eq51}) and (\ref{eq04}) the
respective terms, that appear in addition to the initially mobile fractions
that are still mobile are proportional to each other at all times, as described
in section \ref{chgeneraln}. An analogous relation holds between the immobile
concentration with equilibrium initial conditions and the total
concentration with immobile initial conditions, as can be seen in
equations (\ref{eq51}) and (\ref{eq04}).

\begin{figure}
\centering
\includegraphics{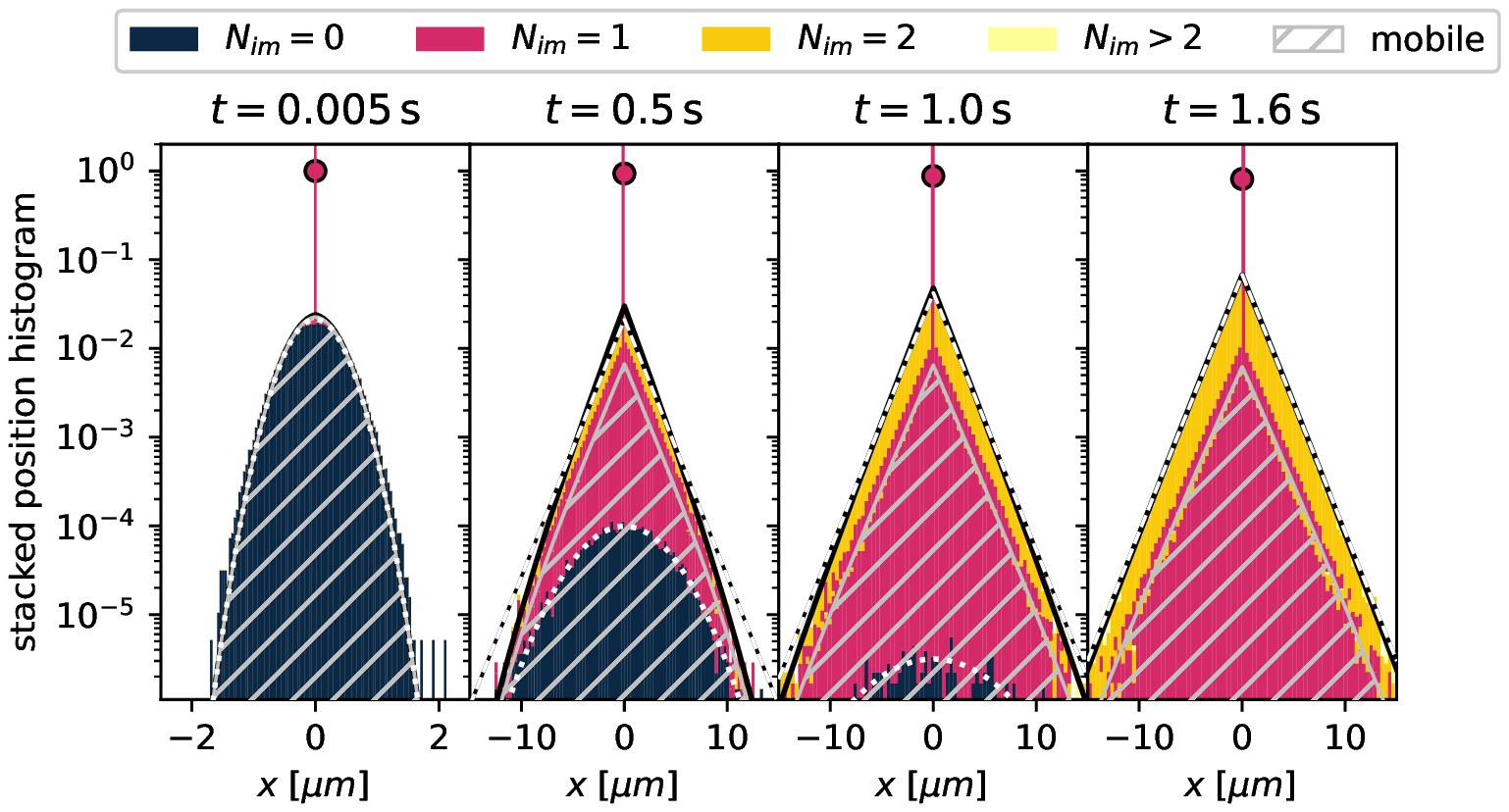}
\includegraphics{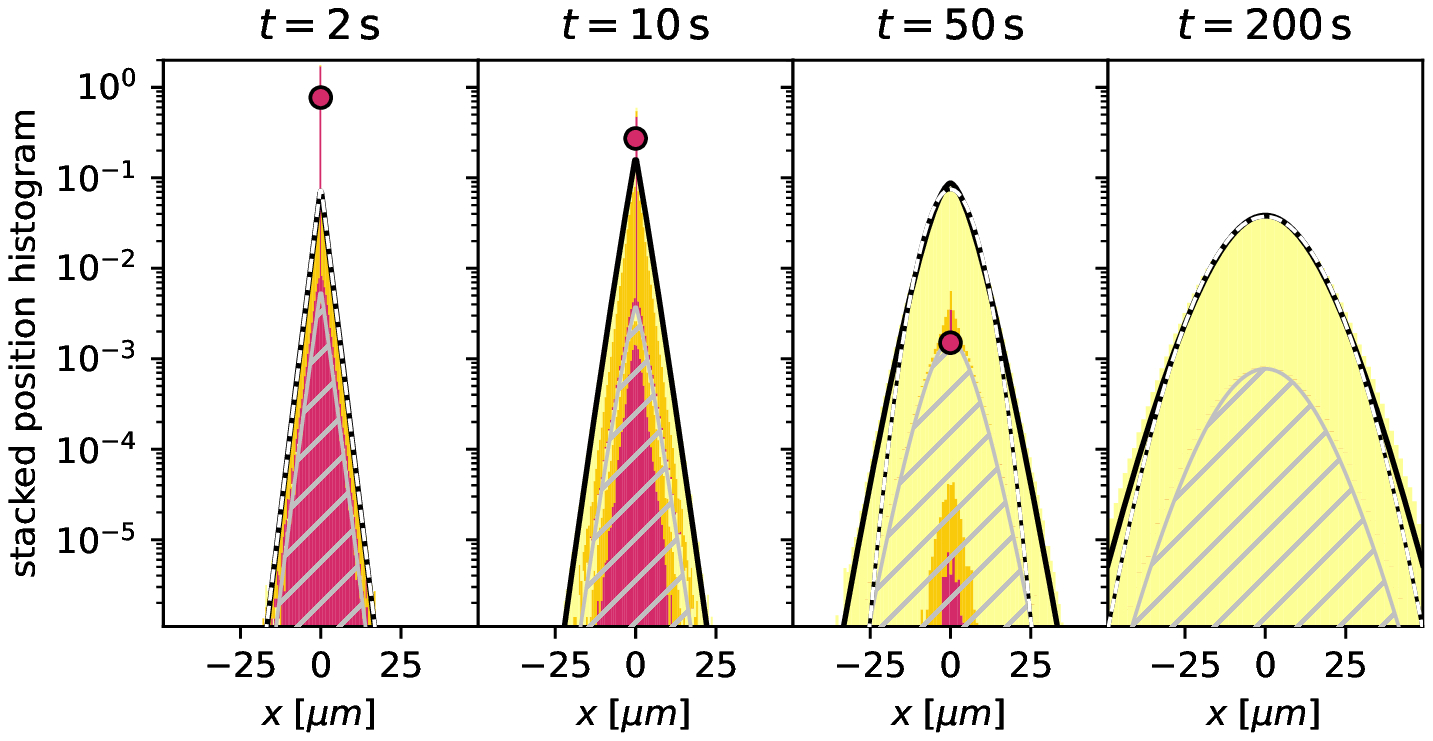}
\caption{Concentration profiles for equilibrium initial conditions. At $t=0$ all
tracers are at $x=0$ and the equilibrium fraction $\tau_\mathrm{im}/(\tau_
\mathrm{m}+\tau_\mathrm{im})$ is immobile. For a description of the legend see
figure \ref{figProfile}. The top
left panel shows the short-time behaviour consisting of a Gaussian and a
$\delta$-distribution, equation (\ref{eq14}). At $t=1$ almost all initially
mobile tracers immobilised at least once and the total concentration follows the
Laplace distribution (\ref{eq007}), as shown by the black-white striped line
for $t=0.5\mathrm{sec}$ to $2\mathrm{sec}$. At
longer times, after several immobilisations the concentration profiles cross
over to a Gaussian, as witnessed by equation (\ref{eq30}), shown as a black-white
striped line for $t=50\mathrm{sec}$ and $t=200\mathrm{sec}$.}
\label{fighisteq}
\end{figure}

We consider the short-time approximation $t\ll\tau_\mathrm{m},\tau_\mathrm{
im}$ for which initially immobile tracers have not yet mobilised and
initially mobile tracers have not yet been trapped. Therefore, we can
neglect the terms with the rates $\tau_{\mathrm{m}}^{-1}$ and $\tau_{\mathrm{
im}}^{-1}$
in (\ref{eq00}) and solve $n_\mathrm{m}(x,t)$ and $n_\mathrm{im}(x,t)$
separately, yielding
\begin{eqnarray} 
\label{eq14}
n_\mathrm{tot}(x,t)\sim\frac{f_\mathrm{m}^\mathrm{eq}}{\sqrt{4\pi Dt}}
\exp\left(-\frac{x^2}{4Dt}\right)+f_\mathrm{im}^\mathrm{eq}\delta(x),
\quad t\ll\tau_\mathrm{m},\tau_\mathrm{im},
\end{eqnarray} 
with a Gaussian distribution describing free diffusion in addition to a
Dirac-$\delta$ distribution of initially immobile tracers that have not
yet moved. This behaviour can be seen in the top left panel of figure
\ref{fighisteq}. The same result as (\ref{eq14}) can be obtained by
combining the short-time expressions for the mobile (\ref{eq44}) and immobile
(\ref{eq45}) initial conditions for $t\ll\tau_\mathrm{m},\tau_\mathrm{im}$,
as done in equation (\ref{eq008}). At short times, the total
density (\ref{eq14}) behaves like the case of mobile initial conditions with an
additional delta peak. At intermediate times $\tau_\mathrm{m}\ll t\ll\tau_
\mathrm{im}$ we obtain
\begin{equation}
\label{eq007}
n_\mathrm{tot}(x,t)\sim\frac{t/\tau_\mathrm{im}}{\sqrt{4D\tau_\mathrm{
m}}}\exp\left(-\frac{|x|}{\sqrt{D\tau_\mathrm{m}}}\right)+\left(1-
\frac{t}{\tau_\mathrm{im}}\right)\delta(x) 
\end{equation}
by combining the mobile (\ref{eq39}) and immobile expression (\ref{eq48}),
respectively\footnote{An approximation for the whole range of
$t\ll\tau_\mathrm{im}$ can be obtained  for any fractions of initially mobile
tracers $f_\mathrm{m}$ by combining equations (\ref{eq50}) and (\ref{eq006})
from the mobile and immobile initial conditions, respectively. This 
yields equation (\ref{eq009}) and is shown in figure
\ref{figntotap}.}. In fact equation (\ref{eq007}) is the same as expression
(\ref{eq48})  for the case of immobile initial conditions, in the intermediate
time regime. This results is shown
in figure \ref{fighisteq} where this approximation is compared to the full
concentration from $t=0.5$ to $t=2$. This result is the one-dimensional
equivalent to the findings in \cite{mora2018brownian}.  The lower right panels
of figure \ref{fighisteq} show the Gaussian long-time limit (\ref{eq30}) as a
black-white striped line.

\subsection{Mean squared displacement}

The number of mobile and immobile tracers remains constant for equilibrium
initial conditions. At all times the second moment of all tracers (\ref{eq03})
thus simplifies to
\begin{equation}
\langle x^2(t)\rangle=\frac{2D}{1+\tau_\mathrm{im}/\tau_\mathrm{m}}t.
\label{eq05}
\end{equation}
The second moment is similar to that of a free Brownian particle, with the
effective diffusion coefficient $D_\mathrm{eff}=D/(1+\tau_\mathrm{im}/
\tau_\mathrm{m})$, as shown in figure \ref{figxx}. This is a known result from
models for Fickian yet non-Gaussian diffusion \cite{mora2018brownian}.  The
mobile and immobile moments, $\langle x^2(t)\rangle_\mathrm{m}$ and $\langle
x^2(t)\rangle_\mathrm{im}$ are equivalent to the moments of the full density with
mobile (\ref{eq07}) and immobile (\ref{eq09}) initial conditions, as can be seen
in figure \ref{figxx}. This relation holds because the respective densities are
proportional, as discussed above.  The mobile and immobile moments show clear
anomalous diffusion for $t\ll\tau_ \mathrm{im}$, with a quite long crossover
dynamics, as depicted in figure \ref{figxx}c. The mobile moment has a plateau in
the intermediate regime $\tau_\mathrm{m}\ll t\ll \tau_\mathrm{im}$ and the
immobile moment behaves ballistically at short times $t\ll \tau_\mathrm{m}$.

In the long-time limit all mobile and immobile second moments grow like the moments
of the total concentration, i.e., $\langle x^2(t)\rangle_\mathrm{m}\sim\langle
x^2(t)\rangle_\mathrm{im}\sim2D_\mathrm{eff}t$ for $t\gg \tau_{\mathrm{im}},
\tau_\mathrm{m}$.

\section{Conclusion}
\label{concl}

We considered a quite simple mobile-immobile model according to which tracer
particles switch between a mobile diffusing state and an immobilised state.
On average, the tracers remain mobile for the duration $\tau_\mathrm{m}$ and
immobile for $\tau_\mathrm{im}$. We considered the particular case, motivated
by experiments on tau proteins binding to and unbinding from microtubules in
axons of dendritic cells \cite{igaev}, when the two timescales are separated
$\tau_\mathrm{m}\ll\tau_\mathrm{im}$. We analysed three different initial
conditions with varying fractions of mobile to immobile tracers at the origin,
that can, in principle, all be realised in experiments.   First, we studied the
case when all tracers are initially mobile, as described in the experiment in
\cite{kreis1982microinjection}. Second,  we assumed all tracers to be initially
immobile. Third, we considered an equilibrium fraction, corresponding to the
experiment in \cite{igaev}. For non-equilibrium fractions of initially mobile
tracers we find anomalous diffusion at short and intermediate timescales, at
which initially mobile tracers display a plateau in the MSD at intermediate
times and initially immobile tracers spread ballistically at short times. At
$t\ll\tau_\mathrm{m}$ and an initial equilibrium fraction, the tracer density
consists of a Gaussian and a delta peak. Initially mobile tracers follow a
Gaussian distribution at short times. When all tracers are initially immobile,
the short time distribution consists of a delta peak and a non-Gaussian
distribution. At intermediate times $\tau_\mathrm{m}\ll t\ll\tau_\mathrm{im}$
the distribution is made up of a Laplace distribution and a delta distribution
of initially immobile tracers that have not moved yet. The coefficients of the
two distributions depend on the specific initial conditions. We additionally
obtain expressions for the densities that are valid for the whole range
$t\ll\tau_\mathrm{im}$.  We stress that the distribution is non-Gaussian at
intermediate times, regardless of the initial conditions. In contrast, the
distribution asymptotically at long times
matches a Gaussian for all initial conditions. The
densities of mobile and immobile tracers with equilibrium initial conditions
match the total tracer densities of mobile and immobile initial conditions,
respectively, at all times. Moreover, the immobile tracer density from  mobile
initial conditions is proportional to the mobile tracer density from immobile
initial conditions at all times.  As a special case for equilibrium initial
conditions, our model corresponds to the one-dimensional version of the model
used in \cite{mora2018brownian} to describe Fickian yet non-Gaussian diffusion.
We find the same linear MSD for all times and obtain a closed expression for the
Laplace distribution at intermediate timescales.

The model developed here is, of course, much more general. We provided the
framework for any ratio of the characteristic time scales $\tau_\mathrm{m}$
and $\tau_\mathrm{im}$, such that the model will be useful for scenarios
ranging from geophysical experiments with Poissonian (im)mobilisation
statistics, to molecular systems such as protein (un)binding to DNA in
nanochannel setups. It will be a topic of future research to study the
effect of a drift velocity in the mobile phase, as well as what happens
when non-exponential (im)mobilisation is considered.

\ack

We acknowledge funding from the German Science Foundation (DFG, grant no.
ME 1535/12-1). AVC acknowledges the support of the Polish National Agency
for Academic Exchange (NAWA).

\appendix

\section{General equations}
\label{secag}

Starting with equation (\ref{eq00}) we apply the Fourier-Laplace transform
$f(k,s)=\int_{-\infty}^\infty \int_0^\infty e^{-s t+ ikx}f(x,t)dtdx$ to the
rate to obtain
\begin{eqnarray}
\nonumber
n_\mathrm{m}(k,s)&=&\frac{f_\mathrm{m}+f_\mathrm{im}\frac{1}{1+s\tau_\mathrm{
im}}}{\phi(s)+k^2D}\\
n_\mathrm{im}(k,s)&=&\frac{\tau_\mathrm{im}}{1+s\tau_\mathrm{im}}\left(f_
\mathrm{im}+\tau_\mathrm{m}^{-1}\frac{f_\mathrm{m}+f_\mathrm{im}\frac{1}{1
+s\tau_\mathrm{im}}}{\phi(s)+k^2D}\right)
\label{eq01}
\end{eqnarray}
as well as
\begin{equation}
\fl n_\mathrm{tot}(k,s)=n_\mathrm{m}(k,s)+n_\mathrm{im}(k,s)=\frac{f_\mathrm{
m}+f_\mathrm{im}\frac{1}{1+s\tau_\mathrm{im}}}{s}\frac{\phi(s)}{\phi(s)+k^2D}
+f_\mathrm{im}\frac{\tau_\mathrm{im}}{1+s\tau_\mathrm{im}}
\label{eq02}
\end{equation}
with $\phi(s)=s[1+\tau_\mathrm{im}\tau_\mathrm{m}^{-1}/(1+s\tau_\mathrm{im})]$.

\begin{figure}
\centering
\includegraphics{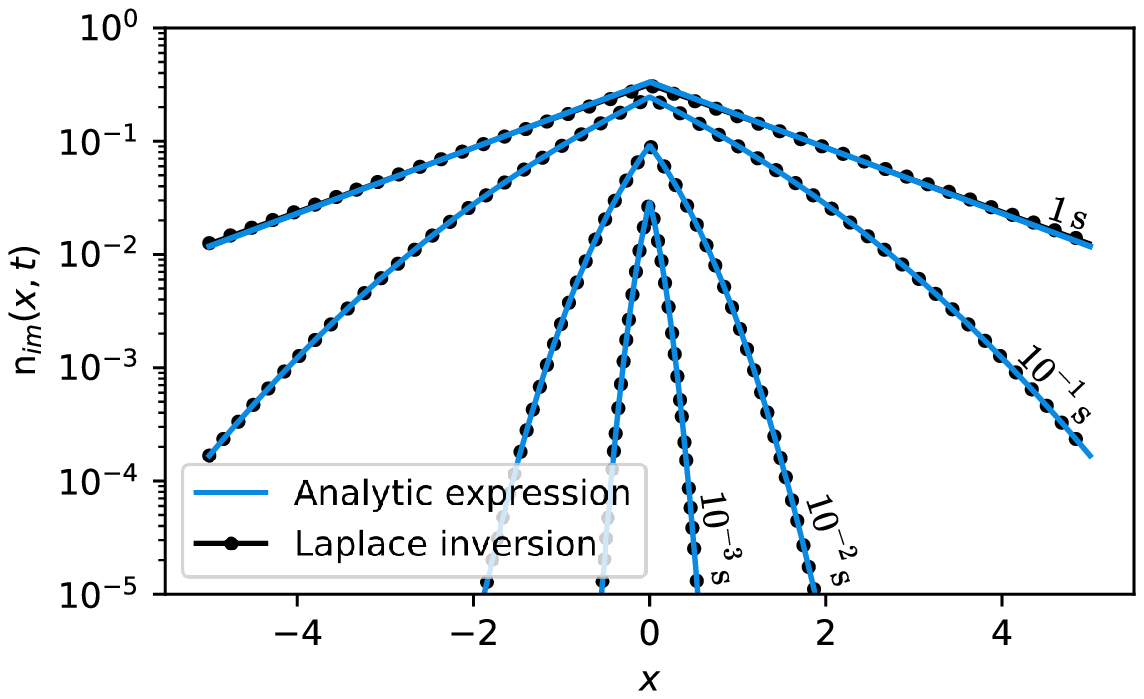}
\caption{Comparison of the Laplace inversion of $n_\mathrm{im}(x,s)$,
expression (\ref{eq04}), and the analytic expression for $n_\mathrm{im}
(x,t)$, equation (\ref{eqer}), that holds for $t\ll\tau_\mathrm{im}$.
Both overlap almost perfectly for $t<\tau_\mathrm{im}=7.7\,\mathrm{sec}$.}
\label{figcimer}
\end{figure}

\section{Asymptotics calculated in Laplace space}

We go from short-time limit to long-time limit.

\subsection{Short-time limit}
\label{chsl}

For $t\ll\tau_\mathrm{m},\tau_\mathrm{im}$, we obtain $s\tau_\mathrm{im}\gg1$
and $s\tau_\mathrm{im}\gg1$. This yields $\phi(s)\sim s$ in this limit.
With (\ref{eq01}) for $f_\mathrm{m}=1$ and $f_\mathrm{im}=0$ we obtain the 
expression
\begin{equation}
n_\mathrm{m}(k,s)\sim n_\mathrm{tot}(k,s)\sim\frac{1}{s+k^2D}
\end{equation}
which produces the Gaussian (\ref{eq44}). We now consider $f_\mathrm{im}=1$
and $f_\mathrm{m}=0$ and obtain the expression 
\begin{equation}
 n_\mathrm{tot}(k,s)\sim
\frac{1}{s\tau_\mathrm{im}}\frac{1}{s+k^2D}
+\left(\frac{1}{s}-\frac{1}{s^2\tau_\mathrm{im}}\right)
\label{eq02a}
\end{equation}
from (\ref{eq02})  in the limit $s\tau_\mathrm{m}\gg 1$ and 
$s\tau_\mathrm{im}\gg 1$. Fourier-Laplace inversion yields the expression
\begin{equation}
\label{eq002}
\fl n_\mathrm{tot}(x,t)\sim \frac{1}{\tau_\mathrm{im}}\int_0^t \frac{1}
{\sqrt{4\pi D t'}}
e^{-\frac{x^2}{4Dt'}}dt'+(1-t/\tau_\mathrm{im})\delta(x)
\quad\mathrm{for}\quad t\ll\tau_\mathrm{m}\ll\tau_\mathrm{im}.
\end{equation}
Solving the integral in (\ref{eq002}) gives the expression
\begin{equation}
\fl n_\mathrm{tot}(x,t)\sim \frac{2t/\tau_\mathrm{im}}{\sqrt{4 \pi D t}}
e^{-\frac{x^2}{4Dt}}
-\frac{|x|(1-\mathrm{erf}\left(\frac{|x|}{\sqrt{4Dt}}\right)}
{2D\tau_\mathrm{im}}
+\left(1-\frac{t}{\tau_\mathrm{im}}\right)\delta(x),\quad\mathrm{for}\quad
t\ll\tau_\mathrm{m}\ll\tau_\mathrm{im},
\label{eq003}
\end{equation}
where normalisation is conserved. By combining expression (\ref{eq003}) for
immobile initial conditions and (\ref{eq44}) for mobile initial conditions, we
obtain the expression
\begin{eqnarray}
\fl n_\mathrm{tot}(x,t)\sim (f_\mathrm{m}+2f_\mathrm{im}t/\tau_\mathrm{im})
\frac{1}{\sqrt{4 \pi D t}}e^{-\frac{x^2}{4Dt}}
-f_\mathrm{im}\frac{|x|\left(1-\mathrm{erf}\left(\frac{|x|}{\sqrt{4Dt}}\right)
\right)}{2D\tau_\mathrm{im}}
+f_\mathrm{im}\left(1-\frac{t}{\tau_\mathrm{im}}\right)\delta(x),\nonumber\\
\label{eq008}
\end{eqnarray}
for $t\ll\tau_\mathrm{m}\ll\tau_\mathrm{im}$ for arbitrary  fractions of
initially mobile tracers.

\subsection{Density at intermediate timescales}
\label{chagi}

We now investigate the intermediate time $\tau_\mathrm{m}\ll t \ll
\tau_\mathrm{im}$, corresponding to $s\tau_\mathrm{m}\ll 1$ and
$s\tau_\mathrm{im}\gg 1$.  In this regime we have $\phi(s)\sim
\tau_\mathrm{m}^{-1}$, yielding the expression 
\begin{eqnarray}
	n_\mathrm{tot}(x,s)\sim
	\frac{f_\mathrm{m}+f_\mathrm{im}\frac{1}{s\tau_\mathrm{im}}}{s}\frac{1}{\sqrt{4D\tau_\mathrm{m}}}e^{-\frac{|x|}{\sqrt{D\tau_\mathrm{m}
	}}} +f_\mathrm{im}
	\left(\frac{1}{s}-\frac{1}{s^2\tau_\mathrm{im}}\right)\delta(x).
		   \label{eq42}
\end{eqnarray}
from (\ref{eq04}) for the total concentration.
Inverse Laplace transform of (\ref{eq42}) yields the expression
\begin{eqnarray}
	n_\mathrm{tot}(x,s)\sim
	(f_\mathrm{m}+f_\mathrm{im} t/\tau_\mathrm{im})\frac{1}
	{\sqrt{4D\tau_\mathrm{m}}}
	e^{-\frac{|x|}{\sqrt{D\tau_\mathrm{m}
	}}} +f_\mathrm{im}
	\left(1-t/\tau_\mathrm{im}\right)\delta(x)
		   \label{eq004}
\end{eqnarray}
for $\tau_\mathrm{m}\ll t\ll\tau_\mathrm{im}$.

\subsection{Density in the long-time limit}
\label{chal}

We obtain the long-time limit
$t\gg\tau_\mathrm{m},\tau_\mathrm{im}$ from $n_\mathrm{tot}(k,s)$ (\ref{eq02})
using $s\ll 1/\tau_\mathrm{im},1/\tau_\mathrm{m}$ and $\phi(s)\sim
s(1+\tau_\mathrm{im}/\tau_\mathrm{im})$. This yields the expression
\begin{equation}
	n_\mathrm{tot}(x,t)\sim \frac{1}{\sqrt{4\pi D_\mathrm{eff} t}}
	e^{-\frac{x^2}{4 D_\mathrm{eff}t}},\quad\mathrm{for}\quad
t\gg\tau_\mathrm{m},\tau_\mathrm{im},
	\label{eq300}
\end{equation}
with $D_\mathrm{eff}=D/(1+\tau_\mathrm{im}/\tau_\mathrm{m})$.

\subsection{Density at short to intermediate timescales}
\label{chImShortInt}

\begin{figure}
\includegraphics[width=\textwidth]{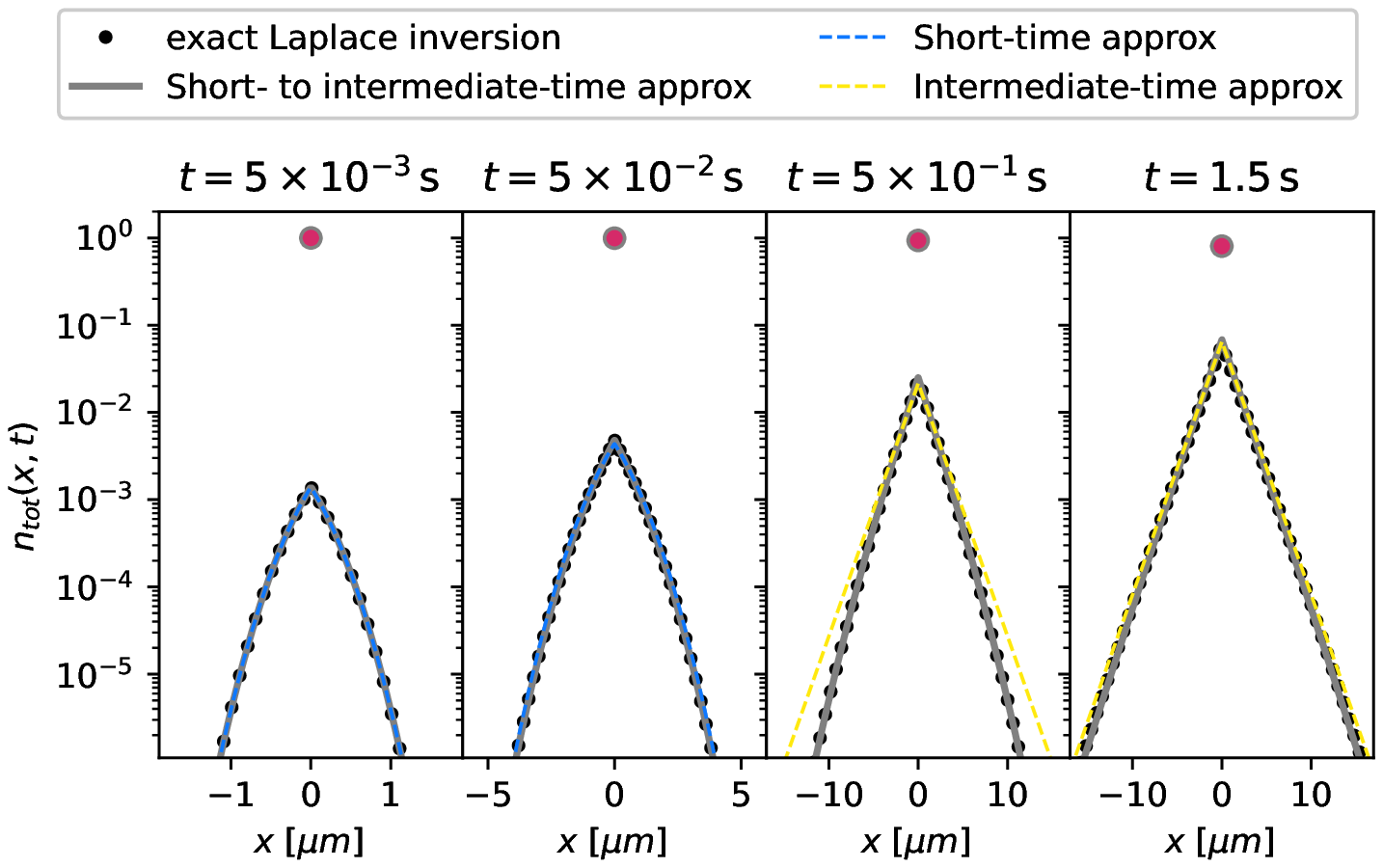}
\caption{All tracers initially immobile.
Comparison of the exact Laplace inversion of (\ref{eq04}),
the short-time approximation (\ref{eq003}),
intermediate time-approximation (\ref{eq004}) and short to intermediate-time
approximation (\ref{eq006}).}
\label{figImmobileApprox}
\end{figure}

\begin{figure}
\centering
\includegraphics{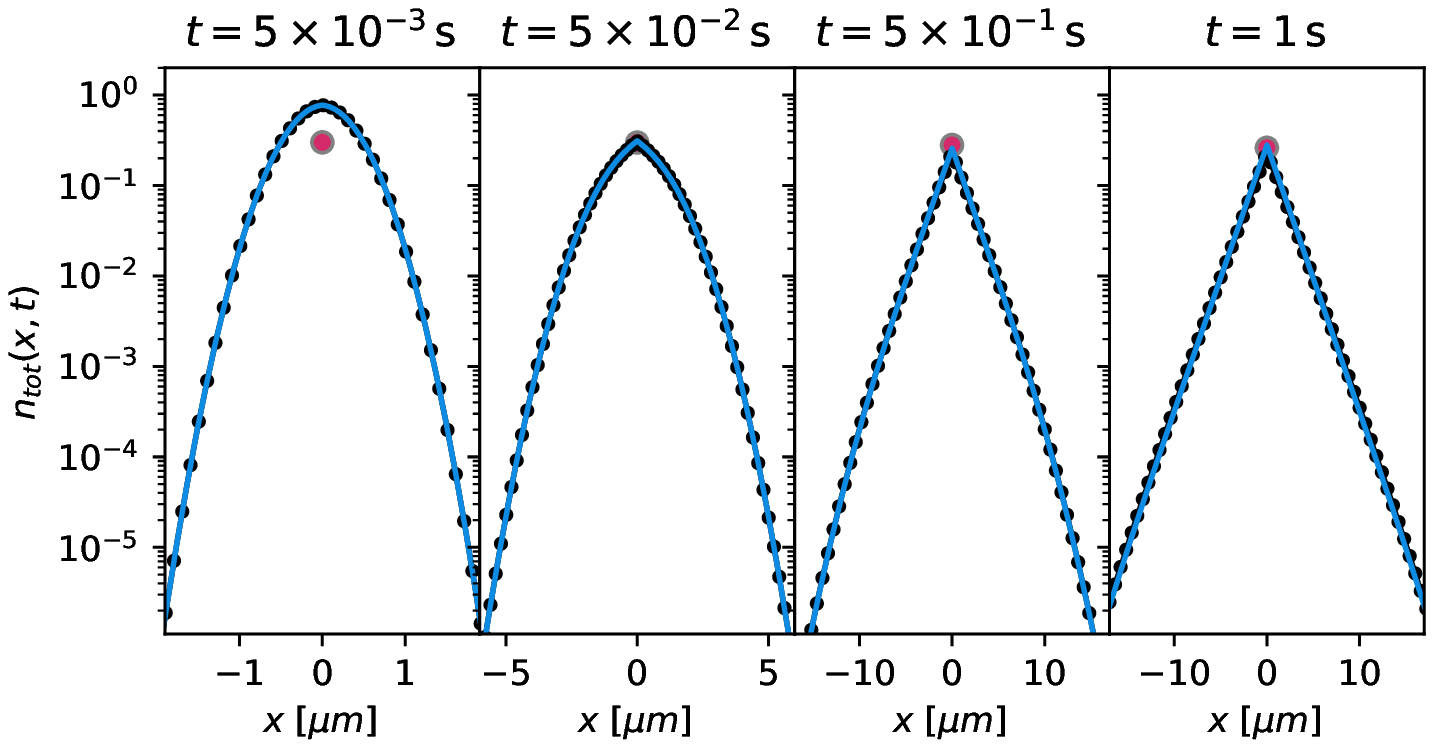}
\caption{Total concentration $n_\mathrm{tot}(x,t)$ for $f_\mathrm{im}=3/10$
and $f_\mathrm{m}=7/10$. Expression (\ref{eq009}) is shown as
blue line and the Laplace inversion of $n_\mathrm{tot}(x,s)$ (\ref{eq04}) is
shown as black line with markers. Both overlap over five decades in amplitude,
for all times shown. The red marker with the grey edge at $x=0$ denotes the
initially immobile tracers that have not yet moved. At short times the
distribution consists of the particles at $x=0$ and a Gaussian. At $t=1s$ the
distribution follows a Laplace distribution (linear tails in the log-linear
plot), on top of the particles at $x=0$.}
\label{figntotap}
\end{figure}

Here we analyse the regime $t\ll\tau_\mathrm{im}$.
The case $f_\mathrm{m}=1$ and
$f_\mathrm{im}=0$ is considered in section \ref{chmobile}. We consider the case
$f_\mathrm{im}=1$ and $f_\mathrm{m}=0$ here.  From $n(x,s)$ (\ref{eq04}), we
obtain with $s\tau_\mathrm{im}\gg 1$ and $\phi(s)\sim s+1/\tau_\mathrm{m}$
\begin{equation}
	\fl n_\mathrm{tot}(x,s)\sim \frac{s\tau_\mathrm{m}+1}
	{s^2\tau_\mathrm{im}\tau_\mathrm{m}}\frac{1}{\sqrt{4D(s+1/\tau_\mathrm{m})}}
	e^{-\sqrt{\frac{s+1/\tau_\mathrm{m}}{D}}|x|}
	+\left(\frac{1}{s}-\frac{1}{s^2\tau_\mathrm{im}}\right),
	\quad\mathrm{for}\quad s\tau_\mathrm{im}\gg 1.
\end{equation}
In time-domain in the limit $t\ll \tau_\mathrm{im}$ this corresponds to 
the expression
\begin{eqnarray}
\label{eq005}
\fl n_\mathrm{tot}(x,t)&\sim&\int_0^t 
\frac{t+\tau_\mathrm{m}-t'}{\tau_\mathrm{im}\tau_\mathrm{m}}
\exp\left(-t'/\tau_\mathrm{m}\right)
\frac{\exp\left(-\frac{x^2}{4Dt'}\right)}{\sqrt{4Dt'}}dt'
+(1-t/\tau_\mathrm{im})\delta(x)\\
&&\hspace*{-2.9cm}=\frac{e^{-\frac{x^2}{4Dt}}}{\sqrt{4\pi D t}}
e^{-t/\tau_\mathrm{m}}\frac{t}{\tau_\mathrm{im}}
+(1-t/\tau_\mathrm{im})\delta(x)\nonumber\\
&&\hspace*{-2.9cm}+\frac{\exp(-|x|/\sqrt{D\tau_\mathrm{m}})}{\sqrt{4D\tau_\mathrm{m}}}
\left(t/\tau_\mathrm{im}-|x|\sqrt{\frac{\tau_m}{4D\tau_\mathrm{im}^2}}
+\tau_\mathrm{m}/2\tau_\mathrm{im}\right)
\frac{1-\mathrm{erf}\left(|x|/\sqrt{4Dt}-\sqrt{t/\tau_\mathrm{m}}\right)}{2}
\nonumber\\
\nonumber
&&\hspace*{-2.9cm}-\frac{\exp(|x|/\sqrt{D\tau_\mathrm{m}})}{\sqrt{4D\tau_\mathrm{m}}}
\left(t/\tau_\mathrm{im}+|x|\sqrt{\frac{\tau_m}{4D\tau_\mathrm{im}^2}}
+\tau_\mathrm{m}/2\tau_\mathrm{im}\right)
\frac{1-\mathrm{erf}\left(|x|/\sqrt{4Dt}+\sqrt{t/\tau_\mathrm{m}}\right)}{2}.\\
\label{eq006}
\end{eqnarray}
Normalisation is preserved, as can be seen by integrating (\ref{eq005}) over $x$.
The first summand in (\ref{eq005}) then resolves to $t/\tau_\mathrm{im}$.
In the limit $t\ll\tau_\mathrm{m},\tau_\mathrm{im}$ we recover
the short-time behaviour for $n_\mathrm{tot}(x,t)$ (\ref{eq003}), as shown in
figure \ref{figImmobileApprox}.
For $\tau_\mathrm{m}\ll t\ll\tau_\mathrm{im}$ and 
$|x|\ll\sqrt{4D\tau_\mathrm{im}^2/\tau_\mathrm{m}}$
we recover the Laplacian intermediate regime in (\ref{eq004})
with $f_\mathrm{im}=1$ and $f_\mathrm{m}=0$.
In figure \ref{figImmobileApprox} we show a verification of (\ref{eq006}).
For arbitrary fractions of initially mobile tracers we combine equation
(\ref{eq006}) for immobile initial conditions with equation (\ref{eq50}) for
mobile initial conditions, as follows
\begin{eqnarray}
n_\mathrm{tot}(x,t)&\sim&	
	\frac{e^{-\frac{x^2}{4Dt}}}{\sqrt{4\pi D t}}
	e^{-t/\tau_\mathrm{m}}\left(f_\mathrm{m}
	+f_\mathrm{im}\frac{t}{\tau_\mathrm{im}}\right)
	+f_\mathrm{im}(1-t/\tau_\mathrm{im})\delta(x)\nonumber\\
	&&+\left[f_\mathrm{m}+f_\mathrm{im}\left(t/\tau_\mathrm{im}
	-|x|\sqrt{\frac{\tau_m}{4D\tau_\mathrm{im}^2}}
	+\tau_\mathrm{m}/2\tau_\mathrm{im}\right)\right]\nonumber\\
	&&\times \frac{\exp(-|x|/\sqrt{D\tau_\mathrm{m}})}{\sqrt{4D\tau_\mathrm{m}}}
	\frac{1-\mathrm{erf}\left(|x|/\sqrt{4Dt}-\sqrt{t/\tau_\mathrm{m}}\right)}{2}
	\nonumber\\
	&&-\left[f_\mathrm{m}+f_\mathrm{im}\left(t/\tau_\mathrm{im}
	+|x|\sqrt{\frac{\tau_m}{4D\tau_\mathrm{im}^2}}
	+\tau_\mathrm{m}/2\tau_\mathrm{im}\right)\right]\nonumber\\
	&&\times \frac{\exp(|x|/\sqrt{D\tau_\mathrm{m}})}{\sqrt{4D\tau_\mathrm{m}}}
	\frac{1-\mathrm{erf}\left(|x|/\sqrt{4Dt}+\sqrt{t/\tau_\mathrm{m}}\right)}{2}.
\label{eq009}
\end{eqnarray}
In figure \ref{figntotap} expression (\ref{eq009}) is compared to the 
Laplace inversion of the exact expression of $n_\mathrm{tot}(x,s)$ (\ref{eq04}).

\end{document}